# An indicator of journal impact that is based on calculating a journal's percentage of highly cited publications

Sara M. González-Betancor [a,*], Pablo Dorta-González [b]

[a] Universidad de Las Palmas de Gran Canaria, Departamento de Métodos Cuantitativos en Economía y Gestión, Facultad de Economía, Empresa y Turismo, Campus de Tafira, 35017 Las Palmas de Gran Canaria, Spain. E-mail: sara.gonzalez@ulpgc.es

[b] Universidad de Las Palmas de Gran Canaria, TiDES Research Institute, Facultad de Economía, Empresa y Turismo, Campus de Tafira, 35017 Las Palmas de Gran Canaria, Spain. E-mail: pablo.dorta@ulpgc.es

[*] Corresponding author and proofs

**Abstract**

The two most used citation impact indicators in the assessment of scientific journals are, nowadays, the impact factor and the h-index. However, both indicators are not field normalized (vary heavily depending on the scientific category) which makes them incomparable between categories. Furthermore, the impact factor is not robust to the presence of articles with a large number of citations, while the h-index depends on the journal size. These limitations are very important when comparing journals of different sizes and categories. An alternative citation impact indicator is the percentage of highly cited articles in a journal. This measure is field normalized (comparable between scientific categories), independent of the journal size and also robust to the presence of articles with a high number of citations. This paper empirically compares this indicator with the impact factor and the h-index, considering different time windows and citation percentiles (levels of citation for considering an article as highly cited compared to others in the same year and category).

## 1. Introduction

The purpose of a research assessment is to evaluate the quality of the research in comparison to other research. As quality is subjective and difficult to measure, citations are used as a proxy. Citations play an important role in scholarly communication and are a significant component in research evaluation. The assumption is that highly cited work has influenced the work of many other researchers and hence it is more valuable.

At present two families of citation impact indicators for scientific journals are often used. The first being the journal impact indices, which consider an average of citations per publication for a given census and citation time window (Garfield, 1972). This family includes, among others, the journal impact factors (for two and five years) of database *Journal Citation Reports* (JCR) owned by Thomson-Reuters (Althouse et al, 2009; Bensman, 2007; Bergstrom, 2007; Bornmann and Daniel, 2008; Moed et al, 2012), the maximum impact factor (Dorta-González and Dorta-González, 2013c), and the *SJR* impact indexes (González-Pereira et al, 2009) of the *Scopus* database owned by Elsevier.

The second of these families of citation impact indicators for scientific journals are the $h$ indices (Hirsch, 2005), which consider that $h$ maximum integer value for which we can say that there are $h$ publications with $h$ or more citations, all within a given time window. This indicator estimates the number of important works published by a journal, increasing the requirement while increasing its value. This is a robust indicator that considers both quantitative and qualitative aspects. However, although this indicator has proven useful for detecting the most prestigious journals in an area, there is empirical evidence which does not discriminate between those at intermediate levels, and penalize selective journals in relation to the major producers (Costas and Bordons, 2007a, b; Dorta-González and Dorta-González, 2011; Egghe, 2013).

Both families of citation impact indicators are useful for comparing journals within the same field. However, they are not appropriate when comparing different scientific fields (Van Raan et al, 2010; Wagner et al, 2011). In this sense there are statistical patterns that allow for the normalization. The average number of references is frequently used in the literature on such normalization (Moed, 2010; Zitt and Small, 2008). However, this reference average is not among the factors that best explain the variability of the indicators (Dorta-González and Dorta-González, 2013a, b).

Traditionally, normalization has also been based on a classification system of journals. This is the case, for example, with the categories in the Web of Science database (Egghe and Rousseau, 2002), the relative position with respect to these categories (Bordons and Barrigón, 1992) and the quartile where each journal belongs when ranked in decreasing order according to their impact factor. However, the delimitation of the scientific fields and disciplines, is a problem not adequately solved in bibliometrics, as these boundaries are diffuse and dynamically developed over time (Leydesdorff, 2012, p. 359). Alternatively, the idea of source normalization has been proposed. In this approach, the normalization is performed according to the citing journals (Dorta-González et al, 2014; Leydesdorff and Bornmann, 2011).

The two families of citation impact indicators most commonly used in the evaluation of journals (impact factors and *h* indices) depend strongly on the scientific field they belong to, which makes them non-comparable across disciplines. In addition, the *h* index also depends on the size of the journal, while the impact factor is not robust with the presence of a small number of highly cited articles. Due to these limitations when comparing journals of different sizes and fields, it is necessary to consider other indicators of impact for journals, to enable comparisons between fields, which do not depend on the size of the journal and be at the same time robust as previously mentioned (Waltman and Van Eck, 2013).

An alternative to this issue is the percentage of highly cited articles in the journal, considering the term article in its strictest sense. Being a percentage, it is a relative value, so this indicator does not depend on the size of the journal. High citation is determined by comparing with the other items in the same field and year at international level, so this indicator does not depend on the field. In addition, it is robust because the inclusion of a new widely cited article does not significantly affect the value of the indicator.

This paper empirically compares this index with the impact factor and the *h* index, considering different time windows and citation percentiles, i.e. levels of citation for considering an article as highly cited compared to others in the same year and field.

## 2. Percentage of highly cited articles in a journal

Highly cited articles are those which have received a number of citations that equals or exceeds the citations of the article that occupies the $q$ percentile position for their category and year of publication. This $q$-value may vary depending on the purpose intended. In this work we have set the following citation percentiles 10%, 20%, 25%, 30%, and 40% as benchmarks.

Having the minimum number of citations necessary to belong to the group of highly cited articles in a category and year of publication, then how many articles meet this requirement in each journal of the category can be determined. Putting this information in relation to the total of articles published that year by the journal (Supplementary material 1), an impact indicator of the journal's scientific production is obtained.

Since the total number of citations of an article is a value that grows over time, it is necessary to set an observation time window. This paper looks at four possible time windows, covering 2, 3, 4, and 5 years.

Let $\left(pArt_q_t\right)_y^j$ be the impact indicator for the $j$-journal and the $y$-year, that measures the percentage of articles in the $q\%$ of the most cited, considering a time window of $t$ years. That is, for a given year and journal, $pArt_q_t$ compares the citations of the articles published in the journal in the period $\left[y-t+1, y\right]$ with those of other articles in the same category and year of the same database, determining what percentage of them go within $q\%$ of the most cited.

For example, $\left(pArt_25_4\right)_{2013}^{JoI}$ identifies the percentage of articles within the first quartile of the most cited, considering a time window of four years, for the Journal of Informetrics (*JoI*) in 2013. This indicator compares the citations of the articles published by *JoI* in the period 2010-2013 with the citations of the rest of articles of the category *Information Science & Library Science* within the Web of Science database during the same period, determining what percentage of them are within 25% of the most cited.

As it is a relative value, the indicator does not depend on the size of the journal. Nor does it depend on the category, as high citation is determined by comparison with other articles within the same category and year. In addition, it is robust because the inclusion of a new widely cited article does not significantly affect the indicator value. This is because all those citations above the ones needed for being highly cited, are not considered for the calculation.

This paper empirically examines this indicator and compares it with the 2-year and 5-year impact factor, and the 3-year and 5-year h-index. Different time windows and benchmarks for a document to be considered highly cited are compared using a total amount of 278 journals and 4 different categories. The percentage of highly cited articles is determined for each journal in terms of five percentage benchmarks (10%, 20%, 25%, 30%, and 40%) and four time windows (2012-13, 2011-2013, 2010-2013, 2009-2013). Thus, twenty indicators (5 benchmarks $\times$ 4 windows) are compared with the 2-year and 5-year impact factor of 2013, and the 3-year and 5-year h-index of 2013. The aim of the study is to identify which of these indicators are field normalized and valid for comparing journals of different categories.

## 3. Materials and Methods

The Thomson Reuters Web of Science classifies journals (approximately 12,000) into 251 scientific categories, which are grouped into 151 research areas and 22 scientific fields. For this research it was decided to work at the level of scientific category and select four categories which are quite different according to the publication profiles and citation levels. Specifically, the categories selected were *Information Science & Library Science (IS & LS)*, *Operations Research & Management Science (OR & MS)*, *Ophthalmology*, and *Physics Condensed Matter (Physics CM)*. Furthermore, it was decided to work with the total population of journals of these categories, instead of a sample.

The information collected from the Thomson Reuters databases for each journal was the following:

- Total number of research articles published between 2009 and 2013
- 5-year h-index (considering the data from 2009 to 2013) and 3-year h-index (considering the data from 2011 to 2013)
- 2-year and 5-year impact factor for 2013

- Number of articles in the journal with enough citations to belong to the group of the 10%, 20%, 25%, 30%, or 40% most cited in its category and year of publication, for each of the five years (2009 to 2013).

With this information we created the database shown in the supplementary material, which is the basis for our research.

It was decided not to include the year 2014, because at the time in which the database searches were made, the impact factor of 2014 had not yet been published. All searches were made between February and March 2015.

## 4. Results

### *4.1. Differences between scientific categories*

Figure 1 shows, based on the information of the Web of Science database, the average number of citations per research article, in the period 2009-2013 for each of the four scientific categories analyzed. This average citation rate is the ratio between the number of citations since publication until 2013 and the number of articles published in each particular year. For example, the value 10.53 in Physics CM in 2011 indicates that, on average, the research articles of 2011 in this category have received a total of 10.53 citations between 2011 and 2013.

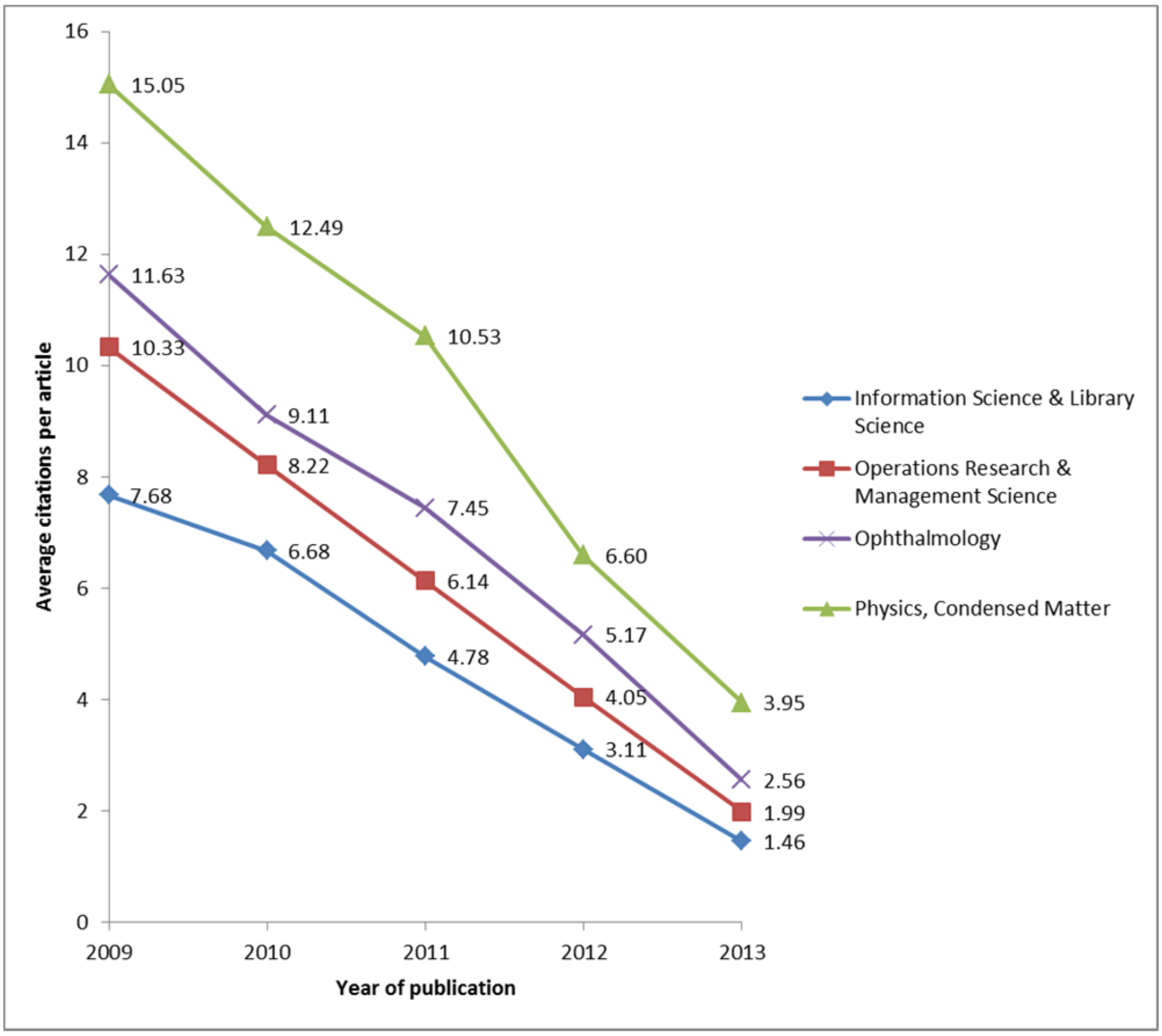

Figure 1: Average citations per article by year of publication. Observe the important differences among journal categories (Source: Thomson Reuters Web of Science).

As shown in Figure 1, the differences in the average number of citations between categories are very important. Thus, the average citations received in Physics CM are about twice those of IS & LS. Ordering the journal categories from the highest to the lowest average citations, the ranking obtained is the following: Physics CM, Ophthalmology, OR & MS, and IS & LS.

The citation percentiles of Figure 2 show the minimum number of citations required to reach a certain percentage benchmark in each journal category and year. For example, in the case of Physics CM, a value of 23 in the 10th percentile for 2011 indicates that 10% of the most cited articles in this category have received at least 23 citations during the period 2011 to 2013.

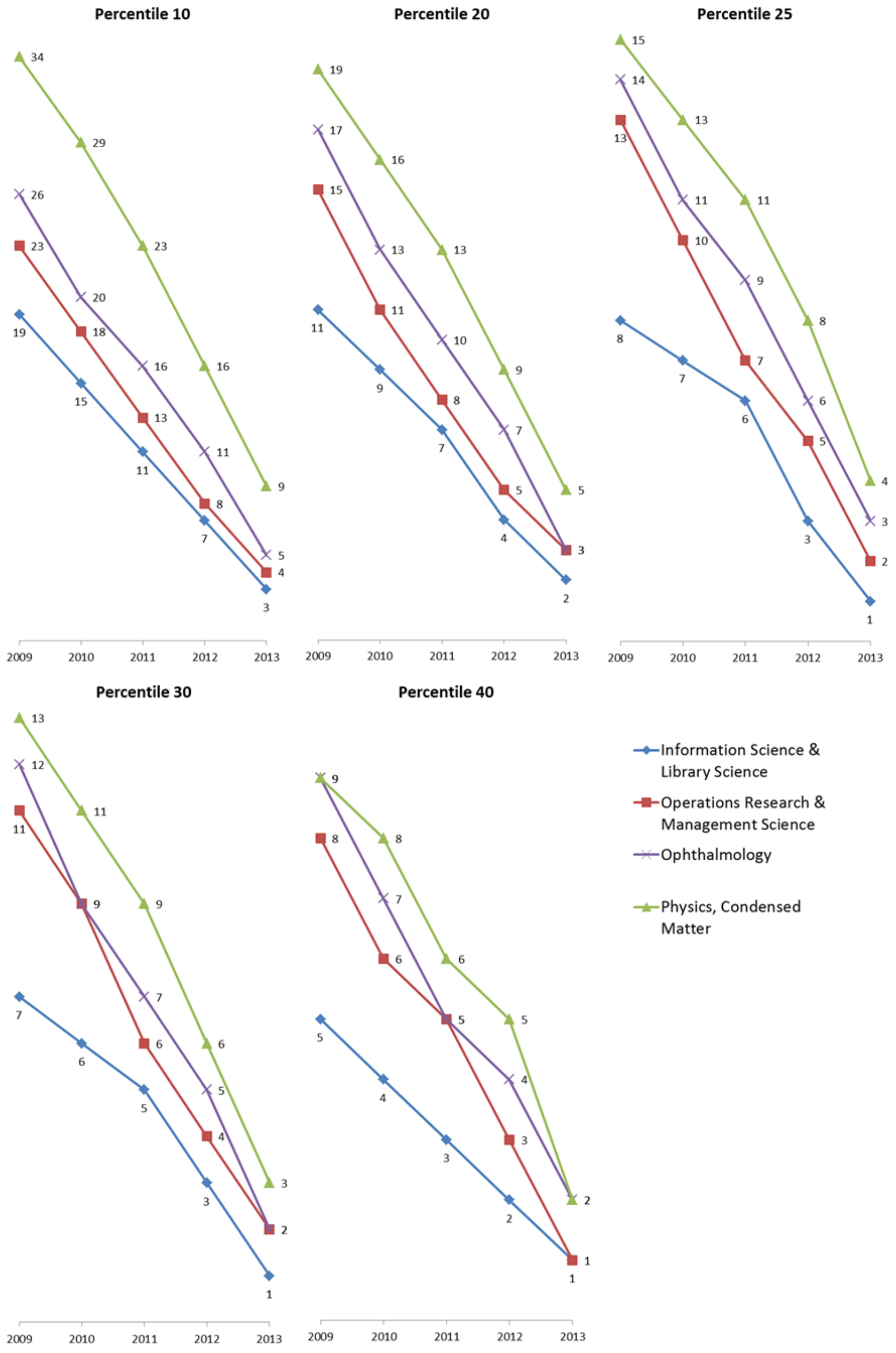

Figure 2: Minimum number of citations necessary to ensure that an article is within 10%, 20%, 25%, 30%, and 40% of the most cited in its journal category and year of publication. Observe again the important differences between journal categories (Source: Thomson Reuters Web of Science).

As can be seen, there are again major differences between categories. Ordering the scientific categories from highest to lowest, according to the required citations needed to reach a specific citation percentile, the same sequence as in the above figure is again obtained: Physics CM, Ophthalmology, OR & MS, and IS & LS.

### *4.2. Field normalization (homogeneity)*

A valid indicator should have the following properties related to homogeneity: it should be homogeneous between scientific categories (between-groups) but heterogeneous within the same scientific category (within-groups). The between-group homogeneity ensures comparability of the indicator between journals of different categories, while the intra-group heterogeneity guarantees the discrimination capacity of the indicator.

Table 1 shows the descriptive statistics of the analyzed variables (*JIF, h,* and *pArt_q_t*) by scientific category. Regarding *pArt*, and within each of the categories, the dispersion of all the proposed indicators is quite large compared to its average (Pearsons' coefficient of variation varies between 0.538 –for *pArt_40_2* in OR & MS– and 1.978 –to *pArt_10_2* in Physics CM–). This is motivated by the important differences between scientific journals in relation to the number of citations received by each of them, and highlights the discrimination capacity of the indicator (intra-group heterogeneity). This relative variability decreases in all scientific categories, as the citation percentile considered grows. Naturally, the same applies when considering all the journals together without differentiating between scientific categories (see Table 2). That is, as the citation percentile decreases, the indicator variability increases, showing that the index is more discriminative.

Table 1: Descriptive statistics of variables (*JIF5*, *JIF2*, *h5*, *h3*, *pArt_q_t*, *q=10, 20, 25, 30, 40; t=5, 4, 3, 2*) in journal categories.

| ***INFORMATION SCIENCE & LIBRARY SCIENCE*** | | | | | | | | |
|---|---|---|---|---|---|---|---|---|
| *Variable* | *Obs* | *Mean* | *Std. Dev.* | *CV* | *Min* | *Max* | *Asymmetry* | *Kurtosis* |
| *JIF5* | 73 | 1.397 | 1.442 | 1.032 | 0.019 | 8.157 | 1.988 | 8.368 |
| *JIF2* | 82 | 1.030 | 0.988 | 0.959 | 0.034 | 5.405 | 1.913 | 7.518 |
| *h5* | 82 | 10.634 | 8.097 | 0.761 | 1 | 39 | 1.486 | 5.256 |
| *h3* | 82 | 6.695 | 4.941 | 0.738 | 1 | 27 | 1.596 | 6.135 |
| *pArt_10_5* | 82 | 0.072 | 0.104 | 1.460 | 0 | 0.556 | 2.201 | 8.700 |
| *pArt_10_4* | 82 | 0.071 | 0.103 | 1.452 | 0 | 0.539 | 2.175 | 8.471 |
| *pArt_10_3* | 82 | 0.071 | 0.104 | 1.460 | 0 | 0.497 | 2.045 | 7.257 |
| *pArt_10_2* | 82 | 0.072 | 0.101 | 1.413 | 0 | 0.470 | 2.008 | 7.202 |
| *pArt_20_5* | 82 | 0.147 | 0.169 | 1.144 | 0 | 0.710 | 1.222 | 3.803 |
| *pArt_20_4* | 82 | 0.147 | 0.168 | 1.144 | 0 | 0.703 | 1.203 | 3.749 |
| *pArt_20_3* | 82 | 0.150 | 0.169 | 1.124 | 0 | 0.675 | 1.146 | 3.476 |
| *pArt_20_2* | 82 | 0.156 | 0.170 | 1.088 | 0 | 0.679 | 1.109 | 3.472 |
| *pArt_25_5* | 82 | 0.221 | 0.206 | 0.933 | 0 | 0.814 | 0.885 | 2.762 |
| *pArt_25_4* | 82 | 0.225 | 0.207 | 0.920 | 0 | 0.819 | 0.855 | 2.716 |
| *pArt_25_3* | 82 | 0.237 | 0.210 | 0.884 | 0 | 0.801 | 0.769 | 2.517 |
| *pArt_25_2* | 82 | 0.275 | 0.221 | 0.806 | 0 | 0.813 | 0.559 | 2.147 |
| *pArt_30_5* | 82 | 0.241 | 0.217 | 0.898 | 0 | 0.824 | 0.768 | 2.476 |
| *pArt_30_4* | 82 | 0.244 | 0.217 | 0.891 | 0 | 0.832 | 0.759 | 2.463 |
| *pArt_30_3* | 82 | 0.249 | 0.215 | 0.864 | 0 | 0.812 | 0.712 | 2.400 |
| *pArt_30_2* | 82 | 0.275 | 0.221 | 0.804 | 0 | 0.813 | 0.560 | 2.150 |
| *pArt_40_5* | 82 | 0.317 | 0.250 | 0.789 | 0 | 0.882 | 0.487 | 1.971 |
| *pArt_40_4* | 82 | 0.320 | 0.250 | 0.781 | 0 | 0.888 | 0.454 | 1.919 |
| *pArt_40_3* | 82 | 0.321 | 0.248 | 0.772 | 0 | 0.874 | 0.443 | 1.931 |
| *pArt_40_2* | 82 | 0.324 | 0.246 | 0.759 | 0 | 0.873 | 0.408 | 1.887 |
| ***OPERATIONS RESEARCH & MANAGEMENT SCIENCE*** | | | | | | | | |
| *Variable* | *Obs* | *Mean* | *Std. Dev.* | *CV* | *Min* | *Max* | *Asymmetry* | *Kurtosis* |
| *JIF5* | 73 | 1.590 | 1.119 | 0.704 | 0.337 | 7.718 | 2.645 | 13.879 |
| *JIF2* | 78 | 1.215 | 0.796 | 0.655 | 0.220 | 4.478 | 1.630 | 6.584 |
| *h5* | 78 | 15.500 | 9.511 | 0.614 | 4 | 52 | 1.578 | 5.793 |
| *h3* | 78 | 9.808 | 5.961 | 0.608 | 2 | 35 | 1.513 | 5.819 |
| *pArt_10_5* | 78 | 0.060 | 0.080 | 1.324 | 0 | 0.495 | 3.031 | 15.041 |
| *pArt_10_4* | 78 | 0.060 | 0.079 | 1.307 | 0 | 0.469 | 2.957 | 14.133 |
| *pArt_10_3* | 78 | 0.061 | 0.079 | 1.286 | 0 | 0.420 | 2.647 | 11.741 |
| *pArt_10_2* | 78 | 0.066 | 0.085 | 1.285 | 0 | 0.457 | 2.417 | 9.965 |
| *pArt_20_5* | 78 | 0.135 | 0.126 | 0.936 | 0 | 0.702 | 1.815 | 7.693 |
| *pArt_20_4* | 78 | 0.136 | 0.125 | 0.917 | 0 | 0.663 | 1.736 | 7.125 |
| *pArt_20_3* | 78 | 0.135 | 0.121 | 0.894 | 0 | 0.596 | 1.585 | 6.230 |
| *pArt_20_2* | 78 | 0.134 | 0.118 | 0.878 | 0 | 0.585 | 1.451 | 5.532 |
| *pArt_25_5* | 78 | 0.174 | 0.145 | 0.831 | 0 | 0.726 | 1.323 | 5.091 |
| *pArt_25_4* | 78 | 0.177 | 0.144 | 0.817 | 0 | 0.691 | 1.265 | 4.784 |
| *pArt_25_3* | 78 | 0.179 | 0.143 | 0.798 | 0 | 0.670 | 1.167 | 4.367 |
| *pArt_25_2* | 78 | 0.183 | 0.141 | 0.770 | 0 | 0.676 | 1.027 | 3.974 |
| *pArt_30_5* | 78 | 0.210 | 0.160 | 0.763 | 0.012 | 0.764 | 1.056 | 3.963 |
| *pArt_30_4* | 78 | 0.211 | 0.159 | 0.750 | 0 | 0.737 | 1.032 | 3.858 |
| *pArt_30_3* | 78 | 0.216 | 0.156 | 0.725 | 0 | 0.701 | 0.895 | 3.401 |
| *pArt_30_2* | 78 | 0.217 | 0.153 | 0.703 | 0 | 0.702 | 0.814 | 3.252 |
| *pArt_40_5* | 78 | 0.318 | 0.190 | 0.597 | 0.024 | 0.851 | 0.515 | 2.569 |
| *pArt_40_4* | 78 | 0.323 | 0.189 | 0.587 | 0.013 | 0.834 | 0.472 | 2.538 |
| *pArt_40_3* | 78 | 0.328 | 0.186 | 0.568 | 0.019 | 0.787 | 0.375 | 2.421 |
| *pArt_40_2* | 78 | 0.350 | 0.189 | 0.538 | 0 | 0.793 | 0.188 | 2.363 |

| ***OPHTHALMOLOGY*** | | | | | | | | |
|---|---|---|---|---|---|---|---|---|
| *Variable* | *Obs* | *Mean* | *Std. Dev.* | *CV* | *Min* | *Max* | *Asymmetry* | *Kurtosis* |
| *JIF5* | 55 | 2.154 | 1.705 | 0.792 | 0.192 | 11.207 | 3.038 | 15.957 |
| *JIF2* | 57 | 2.067 | 1.550 | 0.750 | 0.163 | 9.897 | 2.661 | 13.171 |
| *h5* | 58 | 21.310 | 12.895 | 0.605 | 4 | 61 | 1.062 | 3.606 |
| *h3* | 58 | 14.121 | 8.323 | 0.589 | 3 | 45 | 1.555 | 6.031 |
| *pArt_10_5* | 58 | 0.075 | 0.102 | 1.358 | 0 | 0.701 | 4.170 | 25.564 |
| *pArt_10_4* | 58 | 0.075 | 0.102 | 1.360 | 0 | 0.701 | 4.211 | 25.941 |
| *pArt_10_3* | 58 | 0.076 | 0.104 | 1.366 | 0 | 0.713 | 4.197 | 25.875 |
| *pArt_10_2* | 58 | 0.081 | 0.103 | 1.279 | 0 | 0.707 | 3.979 | 24.185 |
| *pArt_20_5* | 58 | 0.154 | 0.138 | 0.894 | 0 | 0.841 | 2.316 | 11.789 |
| *pArt_20_4* | 58 | 0.156 | 0.138 | 0.886 | 0 | 0.843 | 2.295 | 11.707 |
| *pArt_20_3* | 58 | 0.160 | 0.139 | 0.871 | 0 | 0.851 | 2.259 | 11.585 |
| *pArt_20_2* | 58 | 0.165 | 0.139 | 0.846 | 0 | 0.853 | 2.238 | 11.446 |
| *pArt_25_5* | 58 | 0.180 | 0.149 | 0.826 | 0 | 0.879 | 1.943 | 9.613 |
| *pArt_25_4* | 58 | 0.179 | 0.148 | 0.825 | 0 | 0.881 | 2.002 | 9.979 |
| *pArt_25_3* | 58 | 0.179 | 0.147 | 0.818 | 0 | 0.881 | 2.021 | 10.171 |
| *pArt_25_2* | 58 | 0.181 | 0.144 | 0.796 | 0 | 0.867 | 1.995 | 9.985 |
| *pArt_30_5* | 58 | 0.242 | 0.162 | 0.668 | 0.008 | 0.904 | 1.265 | 6.281 |
| *pArt_30_4* | 58 | 0.246 | 0.163 | 0.662 | 0.010 | 0.910 | 1.237 | 6.186 |
| *pArt_30_3* | 58 | 0.250 | 0.162 | 0.647 | 0.012 | 0.911 | 1.215 | 6.126 |
| *pArt_30_2* | 58 | 0.257 | 0.160 | 0.623 | 0.012 | 0.907 | 1.188 | 5.989 |
| *pArt_40_5* | 58 | 0.299 | 0.176 | 0.587 | 0.008 | 0.924 | 0.767 | 4.381 |
| *pArt_40_4* | 58 | 0.300 | 0.174 | 0.582 | 0.010 | 0.925 | 0.783 | 4.467 |
| *pArt_40_3* | 58 | 0.300 | 0.172 | 0.573 | 0.012 | 0.921 | 0.800 | 4.487 |
| *pArt_40_2* | 58 | 0.288 | 0.165 | 0.573 | 0.012 | 0.907 | 0.896 | 4.828 |
| ***PHYSICS, CONDENSED MATTER*** | | | | | | | | |
| *Variable* | *Obs* | *Mean* | *Std. Dev.* | *CV* | *Min* | *Max* | *Asymmetry* | *Kurtosis* |
| *JIF5* | 59 | 3.296 | 6.216 | 1.886 | 0.098 | 41.775 | 4.492 | 26.468 |
| *JIF2* | 60 | 3.115 | 5.485 | 1.761 | 0.109 | 36.425 | 4.302 | 24.478 |
| *h5* | 60 | 34.667 | 37.272 | 1.075 | 4 | 175 | 2.425 | 8.411 |
| *h3* | 60 | 23.117 | 24.587 | 1.064 | 2 | 112 | 2.313 | 7.696 |
| *pArt_10_5* | 60 | 0.084 | 0.162 | 1.943 | 0 | 0.635 | 2.359 | 7.162 |
| *pArt_10_4* | 60 | 0.082 | 0.160 | 1.957 | 0 | 0.635 | 2.427 | 7.551 |
| *pArt_10_3* | 60 | 0.080 | 0.158 | 1.965 | 0 | 0.635 | 2.461 | 7.711 |
| *pArt_10_2* | 60 | 0.078 | 0.155 | 1.978 | 0 | 0.624 | 2.513 | 7.981 |
| *pArt_20_5* | 60 | 0.160 | 0.217 | 1.355 | 0 | 0.821 | 1.997 | 5.847 |
| *pArt_20_4* | 60 | 0.159 | 0.214 | 1.345 | 0 | 0.821 | 2.004 | 5.943 |
| *pArt_20_3* | 60 | 0.159 | 0.212 | 1.334 | 0 | 0.821 | 2.042 | 6.132 |
| *pArt_20_2* | 60 | 0.160 | 0.209 | 1.309 | 0 | 0.825 | 2.047 | 6.199 |
| *pArt_25_5* | 60 | 0.195 | 0.229 | 1.174 | 0 | 0.866 | 1.799 | 5.270 |
| *pArt_25_4* | 60 | 0.193 | 0.227 | 1.176 | 0 | 0.866 | 1.798 | 5.315 |
| *pArt_25_3* | 60 | 0.192 | 0.224 | 1.169 | 0 | 0.866 | 1.830 | 5.481 |
| *pArt_25_2* | 60 | 0.195 | 0.222 | 1.140 | 0 | 0.863 | 1.816 | 5.467 |
| *pArt_30_5* | 60 | 0.247 | 0.241 | 0.976 | 0.001 | 0.913 | 1.508 | 4.486 |
| *pArt_30_4* | 60 | 0.251 | 0.242 | 0.964 | 0.002 | 0.913 | 1.454 | 4.353 |
| *pArt_30_3* | 60 | 0.255 | 0.240 | 0.940 | 0.002 | 0.913 | 1.432 | 4.345 |
| *pArt_30_2* | 60 | 0.268 | 0.239 | 0.891 | 0 | 0.912 | 1.353 | 4.168 |
| *pArt_40_5* | 60 | 0.337 | 0.251 | 0.743 | 0.007 | 0.948 | 0.939 | 3.327 |
| *pArt_40_4* | 60 | 0.341 | 0.251 | 0.736 | 0.007 | 0.948 | 0.887 | 3.219 |
| *pArt_40_3* | 60 | 0.349 | 0.248 | 0.711 | 0.010 | 0.948 | 0.834 | 3.188 |
| *pArt_40_2* | 60 | 0.355 | 0.245 | 0.690 | 0.014 | 0.943 | 0.840 | 3.219 |

Source: Supplementary material

The average, minimum, and maximum values increase as the citation percentile increases. But just the opposite occurs with the skewness and the kurtosis, that is, these descriptive statistics decrease as the citation percentile increases. This phenomenon is directly related to the definition of the indicator. Note that increasing the citation percentile, the number of articles that meet this condition in every journal increases progressively, reaching the total amount of published articles in the 100-citation-percentile. When this occurs (100 percentile) all indicator values match 1, so that their mean is 1 and their standard deviation is 0. Therefore, as the citation percentile increases, the indicators' distribution tends to become more similar to a standard normal distribution, and hence what we observe about the skewness and kurtosis values.

The minimum values evidence in all categories that, when the citation percentile considered is up to 25, there are journals with no articles among the most cited in its scientific category. In the case of IS & LS that is extensible to at least the 40th percentile.

Moreover, the maximum values evidence the concentration of highly cited articles in some journals. This phenomenon is more common in the Ophthalmology category, where some journals have 70% of their articles among the 10% of the most cited in its category, compared to the 49% maximum in the OR & MS category.

Table 2 shows the descriptive statistics of the analyzed variables for the aggregated data. The dispersion of each of the indicators analyzed, considering all the journals together, is made up of the differences between categories (between-groups) as well as the differences between journals within a category (within-groups). Considering both parts independently, it is observed that the percentage of the variance explained by the differences between groups (categories) is practically zero in all indicators, so that almost 100% is explained by the differences among journals in the same category. This shows that all *pArt* indicators analyzed are homogeneous between groups and heterogeneous within them, so that direct comparison of the indicator value itself between categories is ensured.

Table 2: Descriptive statistics of variables for aggregated data, and analysis of variance (ANOVA)

| *Variable* | *Obs* | *Mean* | *Std. Dev.* | *CV* | *% of variance explained by groups* | *F* | *Prob* | *Min* | *Max* | *Asymmetry* | *Kurtosis* |
|---|---|---|---|---|---|---|---|---|---|---|---|
| *JIF5* | 260 | 2.042 | 3.274 | 1.603 | 0.051 | 4.49 | 0.004 | 0.019 | 41.775 | 7.902 | 87.549 |
| *JIF2* | 277 | 1.747 | 2.836 | 1.623 | 0.094 | 8.14 | 0.000 | 0.034 | 36.425 | 7.889 | 86.477 |
| *h5* | 278 | 19.414 | 21.277 | 1.096 | 0.208 | 19.09 | 0.000 | 1 | 175 | 4.272 | 26.861 |
| *h3* | 278 | 12.662 | 14.035 | 1.108 | 0.226 | 21.09 | 0.000 | 1 | 112 | 4.224 | 25.606 |
| *pArt_10_5* | 278 | 0.072 | 0.113 | 1.577 | 0.000 | 0.51 | 0.678 | 0 | 0.701 | 3.014 | 13.319 |
| *pArt_10_4* | 278 | 0.071 | 0.112 | 1.570 | 0.000 | 0.45 | 0.719 | 0 | 0.701 | 3.047 | 13.646 |
| *pArt_10_3* | 278 | 0.071 | 0.112 | 1.564 | 0.000 | 0.36 | 0.779 | 0 | 0.713 | 2.987 | 13.308 |
| *pArt_10_2* | 278 | 0.073 | 0.111 | 1.512 | 0.000 | 0.24 | 0.869 | 0 | 0.707 | 2.881 | 12.717 |
| *pArt_20_5* | 278 | 0.148 | 0.163 | 1.105 | 0.000 | 0.30 | 0.827 | 0 | 0.841 | 1.943 | 7.265 |
| *pArt_20_4* | 278 | 0.148 | 0.162 | 1.094 | 0.000 | 0.29 | 0.834 | 0 | 0.843 | 1.920 | 7.228 |
| *pArt_20_3* | 278 | 0.150 | 0.161 | 1.077 | 0.000 | 0.35 | 0.788 | 0 | 0.851 | 1.890 | 7.130 |
| *pArt_20_2* | 278 | 0.153 | 0.161 | 1.052 | 0.000 | 0.51 | 0.678 | 0 | 0.853 | 1.856 | 7.036 |
| *pArt_25_5* | 278 | 0.194 | 0.185 | 0.957 | 0.000 | 0.99 | 0.399 | 0 | 0.879 | 1.536 | 5.293 |
| *pArt_25_4* | 278 | 0.195 | 0.185 | 0.949 | 0.002 | 1.13 | 0.337 | 0 | 0.881 | 1.514 | 5.222 |
| *pArt_25_3* | 278 | 0.199 | 0.185 | 0.930 | 0.010 | 1.71 | 0.165 | 0 | 0.881 | 1.468 | 5.040 |
| *pArt_25_2* | 278 | 0.212 | 0.190 | 0.898 | 0.047 | 4.37 | 0.005 | 0 | 0.867 | 1.345 | 4.485 |
| *pArt_30_5* | 278 | 0.234 | 0.197 | 0.843 | 0.000 | 0.54 | 0.654 | 0 | 0.913 | 1.235 | 4.396 |
| *pArt_30_4* | 278 | 0.237 | 0.197 | 0.834 | 0.000 | 0.61 | 0.611 | 0 | 0.913 | 1.209 | 4.328 |
| *pArt_30_3* | 278 | 0.241 | 0.196 | 0.811 | 0.000 | 0.63 | 0.595 | 0 | 0.913 | 1.162 | 4.247 |
| *pArt_30_2* | 278 | 0.253 | 0.197 | 0.777 | 0.005 | 1.33 | 0.265 | 0 | 0.912 | 1.083 | 4.007 |
| *pArt_40_5* | 278 | 0.318 | 0.220 | 0.691 | 0.000 | 0.30 | 0.829 | 0 | 0.948 | 0.712 | 3.002 |
| *pArt_40_4* | 278 | 0.321 | 0.219 | 0.683 | 0.000 | 0.35 | 0.790 | 0 | 0.948 | 0.679 | 2.942 |
| *pArt_40_3* | 278 | 0.325 | 0.217 | 0.668 | 0.000 | 0.49 | 0.687 | 0 | 0.948 | 0.643 | 2.923 |
| *pArt_40_2* | 278 | 0.330 | 0.216 | 0.653 | 0.003 | 1.24 | 0.295 | 0 | 0.943 | 0.595 | 2.864 |

Source: Supplementary material

Table 2 also shows an analysis of variance (ANOVA) for each of the 20 indicators of the *pArt* family, in order to test the null hypothesis that the mean of all categories are equal, versus the alternative that some of them are different (t-test for independent samples). In all cases, except of *pArt_25_2*, there is no statistical evidence to reject the null hypothesis at any level of significance, so that we can speak of homogeneity among groups of indicators. In the case of indicator *pArt_25_2* the category with statistically significant mean difference, compared to other categories, is IS & LS. Since the tests are carried out for the 278 journals together with no missing values in any group, the degrees of freedom are kept constant, so –on the basis that all indicators (except *pArt_25_2*) are equally homogeneous– the best indicator is the one with the highest probability associated to the F statistic, or in other words, the one with the smallest value of F. In this sense, the best indicator is the *pArt_10_2*. It really is the most restrictive of all proposed indicators (percentage of articles within the 10th percentile of citation in a citation window of 2 years).

Furthermore, performing the same hypothesis testing for *JIF* and *h*, it follows that they are not field normalized (Table 2). The tests for *JIF5* and *JIF2* indicate that there are differences between Physics CM and IS & LS, as well as between Physics CM and OR & MS. The tests for *h5* and *h3* indicate that there are differences between Physics CM and the other 3 categories, and between Ophthalmology and IS & LS.

### *4.3. Indicator validity*

In the previous section we have deduced that the *pArt* family of indicators is homogeneous among scientific categories, i.e. it has no differences due to the scientific category of the journal. However, in addition to homogeneity, an indicator must be valid, that is, having the ability to actually measure the impact of the journal. To approach the validity of the *pArt* indicator we have compared the ranking of journals generated within each category using the aforementioned, with those obtained using the most common indicators (*JIF5, JIF2, h5,* and *h3*).

The indicator with the highest level of homogeneity also proved to be the most restrictive –*pArt_10_2*–. In order to assess the validity we decided to analyze the ranking generated through this indicator, and that generated through the least restrictive indicator –*pArt_40_5*–, as it also has the same requirements: homogeneity between-groups (with an associated probability among the highest) and intra-group heterogeneity (Supplementary material 2).

The indicator *pArt_10_2* has null values for 54 of the 278 journals. That is, the number of journals with no articles within the 10th citation percentile, taking a time window of two years, is 54 (26 in IS & LS, 14 in OR & MS, 4 in Ophthalmology, and 10 in Physics CM). Therefore, sorting the journals within each category according to the *pArt_10_2* value, the indicator does not discriminate between journals once the relative positions 57, 65, 55, and 51 are achieved. However, using a higher percentile and time window, such as *pArt_40_5*, the number of journals with nulls is reduced to only 2 cases, specifically in the IS & LS category.

This means that reducing the percentile citation and the time window, makes the indicator less discriminative and it cannot differentiate between lower-impact journals.

Figure 3 provides a comparison between the two proposed indicators and the four most common indicators. The percentage of highly cited articles (*pArt*) show closer, and therefore more comparable, distributions among scientific categories.

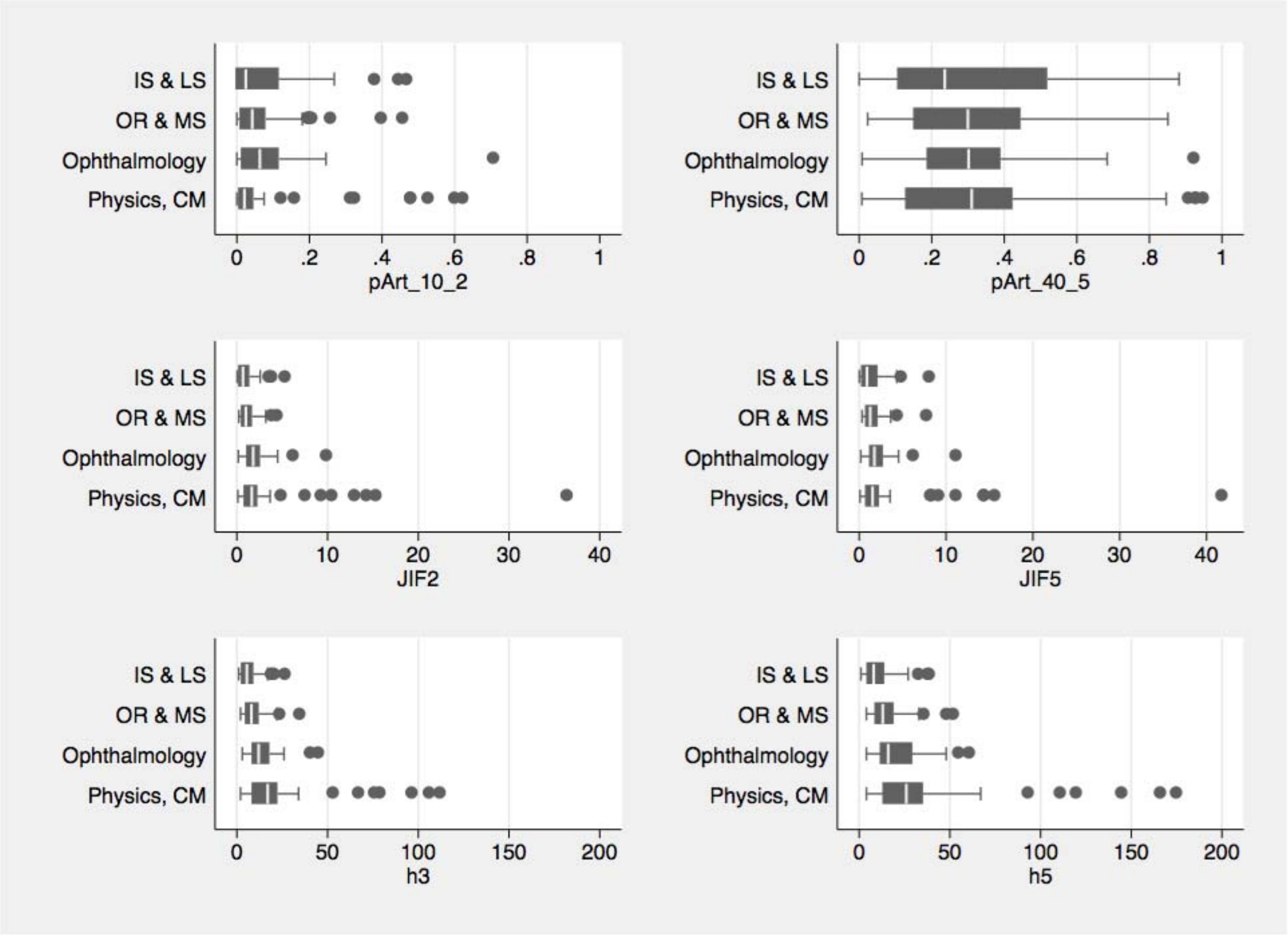

Figure 3: Comparative analysis of indicator distributions among journal categories. Observe that *pArt_40_5* produces distributions closer between categories and therefore is more comparable.

Table 3 shows the Spearman's rank correlation between all analyzed indicators, all of which are significant at 99%. It should be noted that the Spearman's rank correlation is less sensitive than Pearson's to extreme small or large values in the tails of the distribution, as it limits the analysis of such data to their position in the ranking.

Table 3: Spearman's rank correlation between variables in journal categories

| ***INFORMATION SCIENCE & LIBRARY SCIENCE*** | | | | | | |
|---|---|---|---|---|---|---|
| | *JIF2* | *JIF5* | *h3* | *h5* | *pArt_10_2* | *pArt_40_5* |
| *JIF2* | 1 | | | | | |
| *JIF5* | 0.96 | 1 | | | | |
| *h3* | 0.85 | 0.85 | 1 | | | |
| *h5* | 0.88 | 0.91 | 0.96 | 1 | | |
| *pArt_10_2* | 0.84 | 0.83 | 0.88 | 0.85 | 1 | |
| *pArt_40_5* | 0.88 | 0.91 | 0.90 | 0.92 | 0.90 | 1 |
| ***OPERATIONS RESEARCH & MANAGEMENT SCIENCE*** | | | | | | |
| | *JIF2* | *JIF5* | *h3* | *h5* | *pArt_10_2* | *pArt_40_5* |
| *JIF2* | 1 | | | | | |
| *JIF5* | 0.94 | 1 | | | | |
| *h3* | 0.86 | 0.85 | 1 | | | |
| *h5* | 0.83 | 0.88 | 0.95 | 1 | | |
| *pArt_10_2* | 0.68 | 0.69 | 0.64 | 0.65 | 1 | |
| *pArt_40_5* | 0.87 | 0.94 | 0.82 | 0.85 | 0.80 | 1 |
| ***OPHTHALMOLOGY*** | | | | | | |
| | *JIF2* | *JIF5* | *h3* | *h5* | *pArt_10_2* | *pArt_40_5* |
| *JIF2* | 1 | | | | | |
| *JIF5* | 0.95 | 1 | | | | |
| *h3* | 0.84 | 0.81 | 1 | | | |
| *h5* | 0.83 | 0.82 | 0.97 | 1 | | |
| *pArt_10_2* | 0.82 | 0.79 | 0.82 | 0.80 | 1 | |
| *pArt_40_5* | 0.85 | 0.85 | 0.86 | 0.86 | 0.84 | 1 |
| ***PHYSICS, CONDENSED MATTER*** | | | | | | |
| | *JIF2* | *JIF5* | *h3* | *h5* | *pArt_10_2* | *pArt_40_5* |
| *JIF2* | 1 | | | | | |
| *JIF5* | 0.97 | 1 | | | | |
| *h3* | 0.83 | 0.83 | 1 | | | |
| *h5* | 0.83 | 0.85 | 0.98 | 1 | | |
| *pArt_10_2* | 0.93 | 0.91 | 0.82 | 0.82 | 1 | |
| *pArt_40_5* | 0.97 | 0.97 | 0.81 | 0.81 | 0.92 | 1 |

Note: All significant at 99%. Very high correlation between all indicators and therefore very similar journal rankings within each category.

All correlations are quite high and without significant differences between categories. The *pArt_q_t* indicator (percentage of highly cited articles) is robust to parameter changes in *q* and *t*. Indeed, correlations between *pArt_10_2* and *pArt_40_5* are above 0.80 in all categories. Moreover, in general, the correlation between these indicators and *JIF* or *h* are also high, exceeding 0.79, except for OR & MS. It seems, therefore, that the *pArt* indicator provides a similar dimension to that shown by *h* and *JIF*, and also that the *pArt* indicator is robust with respect to any of its parameters.

## 5. Conclusions

The *JIF* and *h* index allow comparisons between journals of the same scientific category, but are not valid indicators to compare journals of different categories. The proportion of highly cited articles in a journal (*pArt*) appears to be an alternative citation impact indicator to achieve this end, since it is a relative measure that lacks the known limitations of other indicators in the literature, such as the journal size or the sensitivity to the high citation of a small number of articles. In fact, it makes it possible to identify what proportion of highly cited articles every journal publishes every year. So that we find journals with 0% of highly cited articles versus journals with more than 90% of highly cited articles, which can be a clear indicator of the impact of a journal.

The dilemma focuses on the limits set to consider an article as part of the group of the most cited, and the time window to be considered in the citation counts. After analyzing the behavior of twenty different indicators, depending on the citation percentile considered (10%, 20%, 25%, 30%, and 40%) and the time window in which the number of citations is counted (2 to 5 years), the indicator that seems to best homogenize the categories is the one that considers a time window of two years and a citation level of 10% (*pArt_10_2*). This indicator is limited by its inability to discriminate within the set of the least cited journals in which none of the articles is within the most cited, due to its restrictions. However, as evidenced in the empirical application, the parameters chosen are not very important, as the results –in terms of homogeneity between categories– have been quite similar in the twenty scenarios studied.

**Acknowledgment**

This research was partially financed by the Ministerio de Economía y Competitividad, grant ECO2014-59067-P.

**SUPPLEMENTARY MATERIAL 1:** Database with the values of the impact indicators

| *id* | *Abbreviated Journal Title* | *WOS Category* | *#Art* | *JIF5* | *JIF2* | *h5* | *h3* | *pArt_ 10_5* | *pArt_ 10_4* | *pArt_ 10_3* | *pArt_ 10_2* | *pArt_ 20_5* | *pArt_ 20_4* | *pArt_ 20_3* | *pArt_ 20_2* | *pArt_ 25_5* | *pArt_ 25_4* | *pArt_ 25_3* | *pArt_ 25_2* | *pArt_ 30_5* | *pArt_ 30_4* | *pArt_ 30_3* | *pArt_ 30_2* | *pArt_ 40_5* | *pArt_ 40_4* | *pArt_ 40_3* | *p 4* |
|---|---|---|---|---|---|---|---|---|---|---|---|---|---|---|---|---|---|---|---|---|---|---|---|---|---|---|---|
| 1 | AFR J LIBR ARCH INFO | 1 | 67 | 0.099 | 0.115 | 2 | 1 | 0.000 | 0.000 | 0.000 | 0.000 | 0.000 | 0.000 | 0.000 | 0.000 | 0.000 | 0.000 | 0.000 | 0.000 | 0.000 | 0.000 | 0.000 | 0.000 | 0.000 | 0.000 | 0.000 | 0 |
| 2 | ASLIB PROC | 1 | 196 | 0.704 | 0.575 | 9 | 5 | 0.015 | 0.006 | 0.009 | 0.000 | 0.087 | 0.070 | 0.077 | 0.077 | 0.168 | 0.158 | 0.197 | 0.244 | 0.194 | 0.171 | 0.205 | 0.244 | 0.291 | 0.247 | 0.291 | 0 |
| 3 | AUST ACAD RES LIBR | 1 | 242 | 0.591 | 0.526 | 6 | 3 | 0.000 | 0.000 | 0.000 | 0.000 | 0.004 | 0.005 | 0.007 | 0.011 | 0.041 | 0.046 | 0.050 | 0.077 | 0.045 | 0.052 | 0.050 | 0.077 | 0.103 | 0.108 | 0.094 | 0 |
| 4 | AUST LIBR J | 1 | 589 |  | 0.226 | 1 | 1 | 0.000 | 0.000 | 0.000 | 0.000 | 0.003 | 0.002 | 0.003 | 0.004 | 0.008 | 0.009 | 0.011 | 0.016 | 0.008 | 0.009 | 0.011 | 0.016 | 0.015 | 0.017 | 0.019 | 0 |
| 5 | CAN J INFORM LIB SCI | 1 | 90 | 0.100 | 0.074 | 3 | 2 | 0.000 | 0.000 | 0.000 | 0.000 | 0.000 | 0.000 | 0.000 | 0.000 | 0.056 | 0.051 | 0.073 | 0.125 | 0.056 | 0.051 | 0.073 | 0.125 | 0.067 | 0.051 | 0.073 | 0 |
| 6 | COLL RES LIBR | 1 | 374 | 1.163 | 1.333 | 9 | 7 | 0.021 | 0.024 | 0.027 | 0.042 | 0.061 | 0.068 | 0.087 | 0.097 | 0.099 | 0.115 | 0.123 | 0.132 | 0.112 | 0.125 | 0.132 | 0.132 | 0.179 | 0.200 | 0.210 | 0 |
| 7 | DATA BASE ADV INF SY | 1 | 104 |  | 0.056 | 6 | 3 | 0.019 | 0.012 | 0.016 | 0.029 | 0.029 | 0.012 | 0.016 | 0.029 | 0.106 | 0.099 | 0.082 | 0.143 | 0.106 | 0.099 | 0.082 | 0.143 | 0.173 | 0.173 | 0.131 | 0 |
| 8 | ECONTENT | 1 | 671 | 0.067 | 0.034 | 2 | 1 | 0.000 | 0.000 | 0.000 | 0.000 | 0.000 | 0.000 | 0.000 | 0.000 | 0.000 | 0.000 | 0.000 | 0.000 | 0.000 | 0.000 | 0.000 | 0.000 | 0.000 | 0.000 | 0.000 | 0 |
| 9 | ELECTRON LIBR | 1 | 477 | 0.381 | 0.228 | 10 | 6 | 0.008 | 0.008 | 0.004 | 0.000 | 0.048 | 0.050 | 0.046 | 0.048 | 0.088 | 0.095 | 0.095 | 0.119 | 0.099 | 0.106 | 0.106 | 0.119 | 0.128 | 0.131 | 0.114 | 0 |
| 10 | ETHICS INF TECHNOL | 1 | 150 |  | 0.520 | 8 | 4 | 0.013 | 0.017 | 0.012 | 0.018 | 0.080 | 0.066 | 0.082 | 0.107 | 0.167 | 0.157 | 0.176 | 0.250 | 0.207 | 0.182 | 0.188 | 0.250 | 0.327 | 0.331 | 0.306 | 0 |
| 11 | EUR J INFORM SYST | 1 | 240 | 2.619 | 1.654 | 15 | 10 | 0.138 | 0.130 | 0.142 | 0.182 | 0.275 | 0.259 | 0.277 | 0.352 | 0.417 | 0.399 | 0.404 | 0.511 | 0.479 | 0.451 | 0.433 | 0.511 | 0.608 | 0.580 | 0.546 | 0 |
| 12 | GOV INFORM Q | 1 | 402 | 2.015 | 2.033 | 19 | 14 | 0.144 | 0.152 | 0.149 | 0.151 | 0.281 | 0.298 | 0.307 | 0.331 | 0.378 | 0.405 | 0.427 | 0.488 | 0.403 | 0.430 | 0.448 | 0.488 | 0.488 | 0.505 | 0.515 | 0 |
| 13 | HEALTH INFO LIBR J | 1 | 220 | 1.093 | 0.932 | 9 | 6 | 0.050 | 0.052 | 0.048 | 0.065 | 0.123 | 0.139 | 0.161 | 0.208 | 0.214 | 0.243 | 0.290 | 0.403 | 0.236 | 0.266 | 0.315 | 0.403 | 0.373 | 0.416 | 0.435 | 0 |
| 14 | INFORM SOC-ESTUD | 1 | 203 | 0.049 | 0.080 | 2 | 1 | 0.000 | 0.000 | 0.000 | 0.000 | 0.000 | 0.000 | 0.000 | 0.000 | 0.005 | 0.006 | 0.008 | 0.012 | 0.005 | 0.006 | 0.008 | 0.012 | 0.005 | 0.006 | 0.008 | 0 |
| 15 | INFORM ORGAN-UK | 1 | 71 | 2.508 | 2.538 | 11 | 9 | 0.183 | 0.236 | 0.295 | 0.242 | 0.296 | 0.309 | 0.364 | 0.333 | 0.451 | 0.491 | 0.591 | 0.636 | 0.535 | 0.600 | 0.636 | 0.636 | 0.634 | 0.673 | 0.705 | 0 |
| 16 | INFORM CULT | 1 | 43 | 0.316 | 0.316 | 2 | 2 | 0.023 | 0.023 | 0.023 | 0.023 | 0.023 | 0.023 | 0.023 | 0.023 | 0.070 | 0.070 | 0.070 | 0.070 | 0.093 | 0.093 | 0.093 | 0.093 | 0.186 | 0.186 | 0.186 | 0 |
| 17 | INFORM DEV | 1 | 171 | 0.357 | 0.440 | 6 | 4 | 0.012 | 0.014 | 0.019 | 0.014 | 0.029 | 0.036 | 0.047 | 0.042 | 0.047 | 0.058 | 0.075 | 0.083 | 0.058 | 0.065 | 0.075 | 0.083 | 0.099 | 0.101 | 0.123 | 0 |
| 18 | INFORM MANAGE-AMSTER | 1 | 240 | 3.392 | 1.788 | 24 | 11 | 0.233 | 0.202 | 0.184 | 0.165 | 0.475 | 0.448 | 0.418 | 0.381 | 0.625 | 0.601 | 0.574 | 0.577 | 0.642 | 0.612 | 0.574 | 0.577 | 0.733 | 0.699 | 0.652 | 0 |
| 19 | INFORM PROCESS MANAG | 1 | 363 | 1.481 | 1.069 | 14 | 9 | 0.099 | 0.086 | 0.103 | 0.115 | 0.236 | 0.234 | 0.273 | 0.299 | 0.390 | 0.380 | 0.417 | 0.494 | 0.407 | 0.393 | 0.417 | 0.494 | 0.522 | 0.515 | 0.533 | 0 |
| 20 | INFORM RES | 1 | 468 | 0.925 | 0.660 | 3 | 2 | 0.000 | 0.000 | 0.000 | 0.000 | 0.002 | 0.000 | 0.000 | 0.000 | 0.009 | 0.005 | 0.007 | 0.009 | 0.011 | 0.005 | 0.007 | 0.009 | 0.015 | 0.010 | 0.013 | 0 |
| 21 | INFORM SOC | 1 | 213 | 1.195 | 0.972 | 10 | 6 | 0.028 | 0.029 | 0.023 | 0.024 | 0.127 | 0.131 | 0.128 | 0.141 | 0.197 | 0.200 | 0.211 | 0.247 | 0.225 | 0.229 | 0.218 | 0.247 | 0.305 | 0.303 | 0.278 | 0 |
| 22 | INFORM SYST J | 1 | 142 | 2.786 | 1.333 | 14 | 7 | 0.113 | 0.106 | 0.105 | 0.153 | 0.324 | 0.319 | 0.349 | 0.373 | 0.458 | 0.442 | 0.430 | 0.475 | 0.479 | 0.469 | 0.430 | 0.475 | 0.641 | 0.637 | 0.616 | 0 |
| 23 | INFORM SYST RES | 1 | 278 | 4.276 | 2.322 | 24 | 12 | 0.302 | 0.259 | 0.241 | 0.236 | 0.504 | 0.474 | 0.471 | 0.479 | 0.626 | 0.607 | 0.592 | 0.614 | 0.644 | 0.623 | 0.602 | 0.614 | 0.745 | 0.729 | 0.723 | 0 |
| 24 | INFORM TECHNOL DEV | 1 | 118 |  | 0.421 | 7 | 3 | 0.008 | 0.011 | 0.014 | 0.000 | 0.042 | 0.021 | 0.029 | 0.021 | 0.136 | 0.106 | 0.145 | 0.191 | 0.153 | 0.128 | 0.159 | 0.191 | 0.237 | 0.202 | 0.246 | 0 |
| 25 | INFORM TECHNOL MANAG | 1 | 122 | 0.942 | 0.897 | 11 | 11 | 0.180 | 0.216 | 0.250 | 0.254 | 0.270 | 0.324 | 0.364 | 0.356 | 0.311 | 0.373 | 0.420 | 0.441 | 0.328 | 0.392 | 0.432 | 0.441 | 0.393 | 0.471 | 0.511 | 0 |
| 26 | INFORM TECHNOL PEOPL | 1 | 102 |  | 0.938 | 7 | 4 | 0.010 | 0.000 | 0.000 | 0.000 | 0.088 | 0.075 | 0.079 | 0.103 | 0.216 | 0.225 | 0.238 | 0.333 | 0.284 | 0.262 | 0.270 | 0.333 | 0.471 | 0.488 | 0.476 | 0 |
| 27 | INTERLEND DOC SUPPLY | 1 | 167 | 0.225 | 0.350 | 5 | 4 | 0.006 | 0.008 | 0.011 | 0.018 | 0.036 | 0.047 | 0.067 | 0.105 | 0.132 | 0.171 | 0.247 | 0.368 | 0.156 | 0.202 | 0.258 | 0.368 | 0.299 | 0.388 | 0.404 | 0 |
| 28 | INT J COMP-SUPP COLL | 1 | 125 | 2.609 | 1.830 | 15 | 10 | 0.232 | 0.196 | 0.200 | 0.163 | 0.376 | 0.343 | 0.387 | 0.347 | 0.576 | 0.559 | 0.587 | 0.612 | 0.600 | 0.578 | 0.587 | 0.612 | 0.736 | 0.725 | 0.725 | 0 |
| 29 | INT J GEOGR INF SCI | 1 | 553 | 1.954 | 1.479 | 20 | 13 | 0.148 | 0.142 | 0.135 | 0.148 | 0.311 | 0.306 | 0.286 | 0.305 | 0.423 | 0.429 | 0.418 | 0.484 | 0.467 | 0.473 | 0.455 | 0.484 | 0.588 | 0.594 | 0.574 | 0 |
| 30 | INT J INFORM MANAGE | 1 | 426 | 2.243 | 2.042 | 21 | 13 | 0.138 | 0.146 | 0.135 | 0.127 | 0.279 | 0.291 | 0.281 | 0.265 | 0.385 | 0.397 | 0.416 | 0.442 | 0.411 | 0.426 | 0.438 | 0.442 | 0.512 | 0.531 | 0.536 | 0 |
| 31 | INVESTIG BIBLIOTECOL | 1 | 619 | 0.133 | 0.062 | 2 | 1 | 0.000 | 0.000 | 0.000 | 0.000 | 0.000 | 0.000 | 0.000 | 0.000 | 0.006 | 0.007 | 0.010 | 0.014 | 0.006 | 0.007 | 0.010 | 0.014 | 0.018 | 0.022 | 0.019 | 0 |
| 32 | J ACAD LIBR | 1 | 601 | 0.838 | 0.574 | 9 | 6 | 0.005 | 0.004 | 0.000 | 0.000 | 0.027 | 0.027 | 0.022 | 0.013 | 0.078 | 0.085 | 0.088 | 0.115 | 0.095 | 0.104 | 0.096 | 0.115 | 0.143 | 0.156 | 0.146 | 0 |
| 33 | J COMPUT-MEDIAT COMM | 1 | 176 | 4.346 | 2.019 | 21 | 10 | 0.239 | 0.241 | 0.253 | 0.242 | 0.460 | 0.457 | 0.482 | 0.452 | 0.585 | 0.586 | 0.627 | 0.629 | 0.625 | 0.621 | 0.639 | 0.629 | 0.744 | 0.716 | 0.747 | 0 |
| 34 | J DOC | 1 | 311 | 1.395 | 1.035 | 12 | 8 | 0.055 | 0.059 | 0.067 | 0.100 | 0.167 | 0.181 | 0.180 | 0.218 | 0.238 | 0.266 | 0.275 | 0.355 | 0.273 | 0.304 | 0.303 | 0.355 | 0.363 | 0.401 | 0.393 | 0 |
| 35 | J GLOB INF MANAG | 1 | 79 | 0.556 | 0.483 | 6 | 3 | 0.000 | 0.000 | 0.000 | 0.000 | 0.013 | 0.000 | 0.000 | 0.000 | 0.076 | 0.063 | 0.042 | 0.061 | 0.127 | 0.094 | 0.042 | 0.061 | 0.215 | 0.172 | 0.104 | 0 |
| 36 | J GLOB INF TECH MAN | 1 | 93 |  | 0.500 | 4 | 3 | 0.011 | 0.011 | 0.000 | 0.000 | 0.032 | 0.032 | 0.028 | 0.042 | 0.065 | 0.065 | 0.056 | 0.083 | 0.065 | 0.065 | 0.056 | 0.083 | 0.118 | 0.118 | 0.111 | 0 |
| 37 | J HEALTH COMMUN | 1 | 595 | 2.355 | 1.869 | 22 | 14 | 0.146 | 0.157 | 0.152 | 0.143 | 0.345 | 0.356 | 0.336 | 0.338 | 0.496 | 0.511 | 0.505 | 0.555 | 0.526 | 0.540 | 0.535 | 0.555 | 0.642 | 0.660 | 0.654 | 0 |
| 38 | J INF SCI | 1 | 254 | 1.387 | 1.087 | 15 | 8 | 0.075 | 0.068 | 0.051 | 0.048 | 0.193 | 0.179 | 0.159 | 0.171 | 0.335 | 0.324 | 0.318 | 0.390 | 0.374 | 0.357 | 0.344 | 0.390 | 0.520 | 0.493 | 0.465 | 0 |
| 39 | J INF TECHNOL | 1 | 163 | 4.917 | 3.789 | 14 | 6 | 0.104 | 0.090 | 0.074 | 0.098 | 0.258 | 0.233 | 0.223 | 0.279 | 0.368 | 0.331 | 0.340 | 0.426 | 0.399 | 0.361 | 0.362 | 0.426 | 0.466 | 0.436 | 0.436 | 0 |
| 40 | J INFORMETR | 1 | 353 | 3.609 | 3.580 | 27 | 17 | 0.391 | 0.391 | 0.395 | 0.381 | 0.595 | 0.596 | 0.605 | 0.608 | 0.703 | 0.707 | 0.714 | 0.729 | 0.728 | 0.726 | 0.730 | 0.729 | 0.819 | 0.817 | 0.827 | 0 |
| 41 | J KNOWL MANAG | 1 | 303 |  | 1.257 | 15 | 10 | 0.116 | 0.114 | 0.129 | 0.124 | 0.330 | 0.345 | 0.339 | 0.345 | 0.488 | 0.524 | 0.550 | 0.611 | 0.538 | 0.563 | 0.585 | 0.611 | 0.677 | 0.699 | 0.696 | 0 |
| 42 | J LIBR INF SCI | 1 | 214 | 0.633 | 0.273 | 7 | 3 | 0.005 | 0.006 | 0.008 | 0.011 | 0.023 | 0.018 | 0.008 | 0.011 | 0.079 | 0.077 | 0.076 | 0.112 | 0.089 | 0.089 | 0.084 | 0.112 | 0.103 | 0.101 | 0.092 | 0 |
| 43 | J MANAGE INFORM SYST | 1 | 226 | 3.305 | 1.925 | 18 | 10 | 0.159 | 0.131 | 0.124 | 0.110 | 0.319 | 0.284 | 0.285 | 0.275 | 0.447 | 0.432 | 0.438 | 0.473 | 0.473 | 0.454 | 0.445 | 0.473 | 0.602 | 0.590 | 0.569 | 0 |
| 44 | J ORGAN END USER COM | 1 | 72 |  | 0.417 | 5 | 3 | 0.014 | 0.014 | 0.019 | 0.029 | 0.014 | 0.014 | 0.019 | 0.029 | 0.111 | 0.111 | 0.111 | 0.176 | 0.139 | 0.139 | 0.130 | 0.176 | 0.208 | 0.208 | 0.148 | 0 |
| 45 | J SCHOLARLY PUBL | 1 | 172 | 0.339 | 0.255 | 4 | 3 | 0.006 | 0.007 | 0.000 | 0.000 | 0.017 | 0.022 | 0.010 | 0.015 | 0.035 | 0.044 | 0.038 | 0.060 | 0.035 | 0.044 | 0.038 | 0.060 | 0.081 | 0.096 | 0.076 | 0 |
| 46 | J STRATEGIC INF SYST | 1 | 127 | 3.130 | 2.571 | 15 | 11 | 0.173 | 0.174 | 0.190 | 0.160 | 0.394 | 0.394 | 0.381 | 0.360 | 0.520 | 0.523 | 0.524 | 0.580 | 0.543 | 0.550 | 0.548 | 0.580 | 0.646 | 0.651 | 0.631 | 0 |
| 47 | J AM MED INFORM ASSN | 1 | 871 | 4.182 | 3.932 | 38 | 27 | 0.393 | 0.402 | 0.418 | 0.445 | 0.582 | 0.587 | 0.591 | 0.642 | 0.684 | 0.697 | 0.699 | 0.775 | 0.712 | 0.719 | 0.715 | 0.775 | 0.798 | 0.799 | 0.796 | 0 |
| 48 | J AM SOC INF SCI TEC | 1 | 1134 | 2.381 | 2.230 | 33 | 19 | 0.161 | 0.162 | 0.161 | 0.152 | 0.284 | 0.288 | 0.297 | 0.283 | 0.399 | 0.406 | 0.436 | 0.471 | 0.425 | 0.429 | 0.453 | 0.471 | 0.519 | 0.513 | 0.558 | 0 |
| 49 | J ASSOC INF SYST | 1 | 167 | 2.795 | 1.250 | 16 | 7 | 0.144 | 0.122 | 0.082 | 0.119 | 0.299 | 0.267 | 0.235 | 0.299 | 0.401 | 0.351 | 0.327 | 0.433 | 0.461 | 0.420 | 0.388 | 0.433 | 0.617 | 0.580 | 0.531 | 0 |
| 50 | J MED LIBR ASSOC | 1 | 422 | 1.080 | 0.979 | 13 | 7 | 0.024 | 0.027 | 0.028 | 0.023 | 0.097 | 0.089 | 0.099 | 0.090 | 0.147 | 0.139 | 0.163 | 0.175 | 0.168 | 0.154 | 0.175 | 0.175 | 0.223 | 0.207 | 0.226 | 0 |
| 51 | KNOWL MAN RES PRACT | 1 | 183 | 0.987 | 0.683 | 8 | 7 | 0.027 | 0.035 | 0.027 | 0.041 | 0.082 | 0.083 | 0.091 | 0.054 | 0.175 | 0.188 | 0.209 | 0.203 | 0.202 | 0.201 | 0.209 | 0.203 | 0.311 | 0.292 | 0.291 | 0 |
| 52 | KNOWL ORGAN | 1 | 192 | 0.485 | 0.448 | 5 | 4 | 0.005 | 0.006 | 0.008 | 0.012 | 0.031 | 0.038 | 0.047 | 0.062 | 0.073 | 0.089 | 0.110 | 0.160 | 0.083 | 0.101 | 0.118 | 0.160 | 0.141 | 0.152 | 0.181 | 0 |
| 53 | LEARN PUBL | 1 | 294 | 0.893 | 1.037 | 10 | 6 | 0.031 | 0.030 | 0.037 | 0.028 | 0.082 | 0.086 | 0.099 | 0.120 | 0.143 | 0.159 | 0.179 | 0.222 | 0.150 | 0.164 | 0.185 | 0.222 | 0.235 | 0.254 | 0.278 | 0 |
| 54 | LIBR INFORM SC | 1 | 52 | 0.224 | 0.312 | 2 | 1 | 0.000 | 0.000 | 0.000 | 0.000 | 0.000 | 0.000 | 0.000 | 0.000 | 0.019 | 0.026 | 0.037 | 0.056 | 0.019 | 0.026 | 0.037 | 0.056 | 0.019 | 0.026 | 0.037 | 0 |
| 55 | LIBR COLLECT ACQUIS | 1 | 120 | 0.500 | 0.276 | 5 | 3 | 0.008 | 0.000 | 0.000 | 0.000 | 0.017 | 0.010 | 0.014 | 0.028 | 0.042 | 0.021 | 0.029 | 0.056 | 0.058 | 0.042 | 0.058 | 0.056 | 0.083 | 0.073 | 0.087 | 0 |

| | | | | | | | | | | | | | | | | | | | | | | | | | | |
|---|---|---|---|---|---|---|---|---|---|---|---|---|---|---|---|---|---|---|---|---|---|---|---|---|---|---|
| 56 | LIBR HI TECH | 1 | 298 | 0.558 | 0.394 | 9 | 6 | 0.013 | 0.013 | 0.013 | 0.021 | 0.057 | 0.048 | 0.060 | 0.063 | 0.121 | 0.123 | 0.167 | 0.208 | 0.128 | 0.132 | 0.173 | 0.208 | 0.188 | 0.193 | 0.247 |
| 57 | LIBR INFORM SCI RES | 1 | 217 | 1.516 | 1.384 | 14 | 8 | 0.074 | 0.068 | 0.053 | 0.035 | 0.198 | 0.192 | 0.205 | 0.186 | 0.309 | 0.316 | 0.348 | 0.395 | 0.336 | 0.345 | 0.364 | 0.395 | 0.498 | 0.503 | 0.508 |
| 58 | LIBR J | 1 | 21887 | 0.292 | 0.237 | 5 | 4 | 0.000 | 0.000 | 0.000 | 0.000 | 0.000 | 0.000 | 0.000 | 0.000 | 0.001 | 0.001 | 0.001 | 0.001 | 0.001 | 0.001 | 0.001 | 0.001 | 0.001 | 0.001 | 0.002 |
| 59 | LIBR QUART | 1 | 194 | 0.711 | 0.861 | 6 | 3 | 0.010 | 0.013 | 0.017 | 0.023 | 0.041 | 0.046 | 0.050 | 0.057 | 0.082 | 0.078 | 0.092 | 0.115 | 0.093 | 0.092 | 0.100 | 0.115 | 0.144 | 0.150 | 0.133 |
| 60 | LIBR RESOUR TECH SER | 1 | 183 | 0.516 | 0.636 | 6 | 4 | 0.000 | 0.000 | 0.000 | 0.000 | 0.033 | 0.034 | 0.026 | 0.036 | 0.066 | 0.061 | 0.052 | 0.072 | 0.066 | 0.061 | 0.052 | 0.072 | 0.131 | 0.136 | 0.139 |
| 61 | LIBR TRENDS | 1 | 223 | 0.454 | 0.262 | 6 | 4 | 0.009 | 0.011 | 0.015 | 0.022 | 0.027 | 0.034 | 0.037 | 0.055 | 0.054 | 0.063 | 0.074 | 0.099 | 0.058 | 0.063 | 0.074 | 0.099 | 0.108 | 0.103 | 0.125 |
| 62 | LIBRI | 1 | 143 | 0.341 | 0.263 | 4 | 3 | 0.000 | 0.000 | 0.000 | 0.000 | 0.000 | 0.000 | 0.000 | 0.000 | 0.035 | 0.042 | 0.057 | 0.086 | 0.035 | 0.042 | 0.057 | 0.086 | 0.098 | 0.110 | 0.102 |
| 63 | MALAYS J LIBR INF SC | 1 | 112 | 0.349 | 0.333 | 5 | 4 | 0.000 | 0.000 | 0.000 | 0.000 | 0.009 | 0.011 | 0.014 | 0.024 | 0.063 | 0.053 | 0.071 | 0.095 | 0.098 | 0.096 | 0.114 | 0.095 | 0.143 | 0.149 | 0.129 |
| 64 | MIS Q EXEC | 1 | 279 | 1.699 | 1.031 | 39 | 21 | 0.556 | 0.539 | 0.497 | 0.470 | 0.710 | 0.703 | 0.675 | 0.679 | 0.814 | 0.819 | 0.801 | 0.813 | 0.824 | 0.832 | 0.812 | 0.813 | 0.882 | 0.888 | 0.874 |
| 65 | MIS QUART | 1 | 101 | 8.157 | 5.405 | 10 | 5 | 0.079 | 0.085 | 0.033 | 0.024 | 0.158 | 0.171 | 0.117 | 0.098 | 0.267 | 0.268 | 0.233 | 0.268 | 0.317 | 0.317 | 0.283 | 0.268 | 0.426 | 0.439 | 0.417 |
| 66 | ONLINE INFORM REV | 1 | 488 | 1.489 | 1.443 | 14 | 9 | 0.037 | 0.037 | 0.037 | 0.025 | 0.100 | 0.090 | 0.095 | 0.084 | 0.158 | 0.151 | 0.163 | 0.168 | 0.176 | 0.164 | 0.166 | 0.168 | 0.246 | 0.233 | 0.217 |
| 67 | PORTAL-LIBR ACAD | 1 | 204 | 0.619 | 0.651 | 7 | 5 | 0.005 | 0.007 | 0.009 | 0.014 | 0.034 | 0.020 | 0.027 | 0.043 | 0.064 | 0.047 | 0.062 | 0.100 | 0.088 | 0.073 | 0.097 | 0.100 | 0.157 | 0.147 | 0.168 |
| 68 | PROF INFORM | 1 | 426 | 0.303 | 0.402 | 8 | 6 | 0.007 | 0.009 | 0.012 | 0.019 | 0.038 | 0.043 | 0.052 | 0.057 | 0.077 | 0.087 | 0.103 | 0.126 | 0.073 | 0.081 | 0.095 | 0.126 | 0.108 | 0.113 | 0.139 |
| 69 | PROGRAM-ELECTRON LIB | 1 | 198 | 0.460 | 0.500 | 6 | 5 | 0.015 | 0.020 | 0.028 | 0.034 | 0.030 | 0.040 | 0.047 | 0.034 | 0.081 | 0.107 | 0.142 | 0.207 | 0.091 | 0.121 | 0.151 | 0.207 | 0.136 | 0.154 | 0.179 |
| 70 | REF USER SERV Q | 1 | 760 | 0.327 | 0.265 | 5 | 4 | 0.000 | 0.000 | 0.000 | 0.000 | 0.003 | 0.003 | 0.004 | 0.003 | 0.008 | 0.008 | 0.009 | 0.010 | 0.012 | 0.012 | 0.011 | 0.010 | 0.022 | 0.023 | 0.020 |
| 71 | RES EVALUAT | 1 | 188 | 1.435 | 1.338 | 12 | 8 | 0.074 | 0.082 | 0.108 | 0.072 | 0.197 | 0.205 | 0.234 | 0.246 | 0.324 | 0.363 | 0.387 | 0.449 | 0.351 | 0.384 | 0.396 | 0.449 | 0.489 | 0.541 | 0.550 |
| 72 | RESTAURATOR | 1 | 89 | 0.333 | 0.484 | 4 | 3 | 0.011 | 0.014 | 0.020 | 0.027 | 0.011 | 0.014 | 0.020 | 0.027 | 0.045 | 0.056 | 0.078 | 0.081 | 0.045 | 0.056 | 0.078 | 0.081 | 0.090 | 0.113 | 0.157 |
| 73 | REV ESP DOC CIENT | 1 | 174 | 0.879 | 0.717 | 7 | 5 | 0.000 | 0.000 | 0.000 | 0.000 | 0.057 | 0.068 | 0.071 | 0.081 | 0.132 | 0.149 | 0.159 | 0.186 | 0.138 | 0.149 | 0.159 | 0.186 | 0.201 | 0.216 | 0.239 |
| 74 | SCIENTIST | 1 | 1414 | 0.195 | 0.351 | 4 | 2 | 0.001 | 0.001 | 0.000 | 0.000 | 0.001 | 0.002 | 0.000 | 0.000 | 0.006 | 0.008 | 0.008 | 0.012 | 0.006 | 0.008 | 0.008 | 0.012 | 0.009 | 0.012 | 0.012 |
| 75 | SCIENTOMETRICS | 1 | 1180 | 2.294 | 2.274 | 27 | 19 | 0.217 | 0.233 | 0.250 | 0.268 | 0.412 | 0.430 | 0.444 | 0.457 | 0.561 | 0.584 | 0.608 | 0.656 | 0.600 | 0.618 | 0.636 | 0.656 | 0.714 | 0.736 | 0.740 |
| 76 | SERIALS REV | 1 | 262 | 0.618 | 0.531 | 6 | 5 | 0.011 | 0.014 | 0.018 | 0.025 | 0.027 | 0.033 | 0.030 | 0.034 | 0.076 | 0.095 | 0.107 | 0.134 | 0.084 | 0.100 | 0.112 | 0.134 | 0.122 | 0.133 | 0.142 |
| 77 | SOC SCI COMPUT REV | 1 | 198 | 1.686 | 1.542 | 14 | 9 | 0.101 | 0.101 | 0.095 | 0.079 | 0.308 | 0.335 | 0.349 | 0.371 | 0.409 | 0.449 | 0.484 | 0.551 | 0.439 | 0.475 | 0.500 | 0.551 | 0.535 | 0.563 | 0.571 |
| 78 | SOC SCI INFORM | 1 | 157 | 0.604 | 0.594 | 6 | 5 | 0.025 | 0.031 | 0.040 | 0.048 | 0.051 | 0.063 | 0.079 | 0.111 | 0.178 | 0.211 | 0.257 | 0.365 | 0.217 | 0.234 | 0.277 | 0.365 | 0.312 | 0.328 | 0.347 |
| 79 | TELECOMMUN POLICY | 1 | 404 | 1.506 | 1.128 | 16 | 10 | 0.097 | 0.102 | 0.095 | 0.107 | 0.208 | 0.201 | 0.187 | 0.171 | 0.332 | 0.323 | 0.322 | 0.342 | 0.354 | 0.343 | 0.333 | 0.342 | 0.493 | 0.483 | 0.473 |
| 80 | TELEMAT INFORM | 1 | 106 | | 0.705 | 7 | 7 | 0.104 | 0.104 | 0.104 | 0.114 | 0.292 | 0.292 | 0.292 | 0.329 | 0.349 | 0.349 | 0.349 | 0.405 | 0.377 | 0.377 | 0.377 | 0.405 | 0.500 | 0.500 | 0.500 |
| 81 | TRANSINFORMACAO | 1 | 100 | 0.074 | 0.083 | 2 | 2 | 0.000 | 0.000 | 0.000 | 0.000 | 0.000 | 0.000 | 0.000 | 0.000 | 0.010 | 0.013 | 0.017 | 0.024 | 0.010 | 0.013 | 0.017 | 0.024 | 0.030 | 0.038 | 0.050 |
| 82 | Z BIBL BIBL | 1 | 290 | 0.019 | 0.049 | 2 | 2 | 0.003 | 0.005 | 0.006 | 0.009 | 0.003 | 0.005 | 0.006 | 0.009 | 0.007 | 0.009 | 0.012 | 0.019 | 0.007 | 0.009 | 0.012 | 0.019 | 0.010 | 0.014 | 0.018 |
| 83 | 4OR-Q J OPER RES | 2 | 162 | 1.181 | 0.918 | 10 | 5 | 0.000 | 0.000 | 0.000 | 0.000 | 0.068 | 0.085 | 0.082 | 0.063 | 0.123 | 0.140 | 0.143 | 0.156 | 0.142 | 0.163 | 0.163 | 0.172 | 0.216 | 0.240 | 0.245 |
| 84 | ANN OPER RES | 2 | 851 | 1.312 | 1.103 | 18 | 10 | 0.049 | 0.055 | 0.050 | 0.058 | 0.114 | 0.126 | 0.124 | 0.144 | 0.170 | 0.187 | 0.192 | 0.230 | 0.207 | 0.223 | 0.233 | 0.276 | 0.327 | 0.343 | 0.348 |
| 85 | APPL STOCH MODEL BUS | 2 | 282 | 0.751 | 0.532 | 10 | 6 | 0.011 | 0.013 | 0.011 | 0.019 | 0.057 | 0.057 | 0.057 | 0.074 | 0.060 | 0.062 | 0.063 | 0.083 | 0.082 | 0.084 | 0.091 | 0.111 | 0.149 | 0.150 | 0.154 |
| 86 | ASIA PAC J OPER RES | 2 | 243 | 0.396 | 0.220 | 7 | 5 | 0.004 | 0.005 | 0.000 | 0.000 | 0.012 | 0.010 | 0.007 | 0.009 | 0.021 | 0.020 | 0.020 | 0.019 | 0.037 | 0.035 | 0.039 | 0.019 | 0.123 | 0.125 | 0.145 |
| 87 | CENT EUR J OPER RES | 2 | 203 | 0.842 | 0.787 | 9 | 6 | 0.025 | 0.029 | 0.022 | 0.031 | 0.079 | 0.092 | 0.096 | 0.112 | 0.123 | 0.138 | 0.147 | 0.173 | 0.133 | 0.149 | 0.162 | 0.184 | 0.246 | 0.276 | 0.316 |
| 88 | COMPUT OPTIM APPL | 2 | 459 | 1.355 | 0.977 | 16 | 10 | 0.054 | 0.056 | 0.054 | 0.057 | 0.172 | 0.179 | 0.169 | 0.175 | 0.218 | 0.224 | 0.213 | 0.232 | 0.264 | 0.263 | 0.248 | 0.272 | 0.401 | 0.406 | 0.382 |
| 89 | COMPUT OPER RES | 2 | 1282 | 2.335 | 1.718 | 36 | 21 | 0.190 | 0.185 | 0.183 | 0.178 | 0.341 | 0.339 | 0.327 | 0.314 | 0.406 | 0.403 | 0.394 | 0.384 | 0.464 | 0.454 | 0.445 | 0.432 | 0.611 | 0.604 | 0.597 |
| 90 | CONCURRENT ENG-RES A | 2 | 121 | 0.672 | 0.531 | 7 | 5 | 0.008 | 0.010 | 0.014 | 0.020 | 0.025 | 0.031 | 0.041 | 0.061 | 0.058 | 0.072 | 0.095 | 0.143 | 0.058 | 0.072 | 0.095 | 0.143 | 0.223 | 0.258 | 0.284 |
| 91 | DECIS SUPPORT SYST | 2 | 820 | 2.651 | 2.036 | 29 | 19 | 0.161 | 0.158 | 0.164 | 0.149 | 0.321 | 0.329 | 0.339 | 0.303 | 0.389 | 0.403 | 0.417 | 0.382 | 0.441 | 0.452 | 0.461 | 0.431 | 0.582 | 0.604 | 0.613 |
| 92 | DISCRETE EVENT DYN S | 2 | 104 | 1.010 | 0.667 | 7 | 5 | 0.019 | 0.024 | 0.016 | 0.025 | 0.087 | 0.085 | 0.097 | 0.125 | 0.154 | 0.159 | 0.177 | 0.250 | 0.163 | 0.171 | 0.194 | 0.275 | 0.250 | 0.280 | 0.290 |
| 93 | DISCRETE OPTIM | 2 | 165 | 0.917 | 0.629 | 10 | 7 | 0.018 | 0.016 | 0.020 | 0.018 | 0.073 | 0.087 | 0.099 | 0.071 | 0.103 | 0.111 | 0.129 | 0.089 | 0.127 | 0.135 | 0.158 | 0.107 | 0.242 | 0.262 | 0.297 |
| 94 | ENG ECON | 2 | 92 | | 0.647 | 5 | 4 | 0.022 | 0.026 | 0.034 | 0.051 | 0.033 | 0.039 | 0.034 | 0.051 | 0.033 | 0.039 | 0.034 | 0.051 | 0.054 | 0.066 | 0.052 | 0.051 | 0.109 | 0.132 | 0.121 |
| 95 | ENG OPTIMIZ | 2 | 355 | 1.370 | 1.230 | 14 | 10 | 0.048 | 0.052 | 0.061 | 0.050 | 0.149 | 0.166 | 0.189 | 0.170 | 0.189 | 0.210 | 0.241 | 0.233 | 0.228 | 0.248 | 0.285 | 0.264 | 0.363 | 0.372 | 0.404 |
| 96 | EUR J IND ENG | 2 | 146 | 1.554 | 1.500 | 11 | 7 | 0.027 | 0.033 | 0.031 | 0.027 | 0.130 | 0.139 | 0.122 | 0.110 | 0.158 | 0.164 | 0.153 | 0.151 | 0.185 | 0.180 | 0.173 | 0.164 | 0.342 | 0.336 | 0.306 |
| 97 | EUR J OPER RES | 2 | 3072 | 2.625 | 1.843 | 48 | 24 | 0.164 | 0.164 | 0.169 | 0.180 | 0.319 | 0.326 | 0.325 | 0.329 | 0.379 | 0.384 | 0.391 | 0.397 | 0.427 | 0.431 | 0.442 | 0.443 | 0.576 | 0.588 | 0.596 |
| 98 | EXPERT SYST APPL | 2 | 6295 | 2.254 | 1.965 | 52 | 35 | 0.195 | 0.199 | 0.209 | 0.259 | 0.296 | 0.284 | 0.272 | 0.239 | 0.389 | 0.390 | 0.397 | 0.417 | 0.449 | 0.449 | 0.459 | 0.479 | 0.569 | 0.565 | 0.568 |
| 99 | FLEX SERV MANUF J | 2 | 93 | 1.180 | 1.439 | 7 | 6 | 0.054 | 0.057 | 0.067 | 0.091 | 0.097 | 0.103 | 0.120 | 0.145 | 0.118 | 0.126 | 0.147 | 0.164 | 0.151 | 0.161 | 0.187 | 0.218 | 0.269 | 0.287 | 0.293 |
| 100 | FUZZY OPTIM DECIS MA | 2 | 114 | 2.055 | 1.000 | 12 | 8 | 0.158 | 0.141 | 0.153 | 0.196 | 0.281 | 0.272 | 0.319 | 0.333 | 0.395 | 0.402 | 0.458 | 0.471 | 0.439 | 0.446 | 0.514 | 0.529 | 0.561 | 0.565 | 0.611 |
| 101 | IEEE SYST J | 2 | 325 | 1.753 | 1.746 | 16 | 10 | 0.117 | 0.129 | 0.127 | 0.173 | 0.203 | 0.224 | 0.211 | 0.263 | 0.249 | 0.279 | 0.277 | 0.327 | 0.292 | 0.324 | 0.324 | 0.359 | 0.415 | 0.460 | 0.465 |
| 102 | IIE TRANS | 2 | 390 | 1.627 | 1.064 | 16 | 9 | 0.072 | 0.077 | 0.081 | 0.088 | 0.149 | 0.163 | 0.162 | 0.165 | 0.205 | 0.223 | 0.235 | 0.235 | 0.259 | 0.267 | 0.282 | 0.271 | 0.405 | 0.433 | 0.453 |
| 103 | IMA J MANAG MATH | 2 | 139 | 0.688 | 0.471 | 7 | 4 | 0.007 | 0.009 | 0.012 | 0.018 | 0.029 | 0.027 | 0.024 | 0.036 | 0.050 | 0.054 | 0.060 | 0.073 | 0.086 | 0.090 | 0.084 | 0.091 | 0.173 | 0.189 | 0.205 |
| 104 | INFOR | 2 | 119 | 0.465 | 0.410 | 6 | 4 | 0.000 | 0.000 | 0.000 | 0.000 | 0.025 | 0.033 | 0.031 | 0.000 | 0.034 | 0.044 | 0.046 | 0.000 | 0.034 | 0.044 | 0.046 | 0.000 | 0.059 | 0.067 | 0.077 |
| 105 | INFORMS J COMPUT | 2 | 254 | 1.722 | 1.120 | 16 | 8 | 0.075 | 0.077 | 0.065 | 0.075 | 0.157 | 0.155 | 0.148 | 0.140 | 0.205 | 0.193 | 0.194 | 0.196 | 0.244 | 0.227 | 0.219 | 0.224 | 0.378 | 0.372 | 0.368 |
| 106 | INTERFACES | 2 | 371 | 0.669 | 0.443 | 10 | 5 | 0.008 | 0.007 | 0.004 | 0.006 | 0.027 | 0.024 | 0.022 | 0.012 | 0.030 | 0.028 | 0.026 | 0.018 | 0.046 | 0.045 | 0.044 | 0.036 | 0.127 | 0.135 | 0.140 |
| 107 | INT J COMPUT INTEG M | 2 | 415 | 1.143 | 1.019 | 13 | 9 | 0.024 | 0.024 | 0.027 | 0.029 | 0.106 | 0.107 | 0.114 | 0.106 | 0.147 | 0.155 | 0.173 | 0.171 | 0.181 | 0.188 | 0.208 | 0.200 | 0.318 | 0.334 | 0.357 |
| 108 | INT J INF TECH DECIS | 2 | 255 | 1.688 | 1.890 | 17 | 12 | 0.086 | 0.074 | 0.084 | 0.063 | 0.173 | 0.163 | 0.157 | 0.099 | 0.192 | 0.181 | 0.181 | 0.126 | 0.247 | 0.223 | 0.205 | 0.144 | 0.329 | 0.302 | 0.271 |
| 109 | INT J PROD RES | 2 | 2038 | 1.718 | 1.323 | 24 | 17 | 0.078 | 0.077 | 0.083 | 0.093 | 0.195 | 0.202 | 0.198 | 0.197 | 0.257 | 0.267 | 0.267 | 0.272 | 0.300 | 0.308 | 0.310 | 0.304 | 0.466 | 0.476 | 0.473 |
| 110 | INT J SYST SCI | 2 | 817 | 1.714 | 1.579 | 19 | 15 | 0.109 | 0.123 | 0.136 | 0.159 | 0.235 | 0.258 | 0.277 | 0.299 | 0.300 | 0.330 | 0.359 | 0.391 | 0.345 | 0.367 | 0.402 | 0.442 | 0.490 | 0.512 | 0.545 |
| 111 | INT J TECHNOL MANAGE | 2 | 374 | 0.659 | 0.492 | 9 | 5 | 0.005 | 0.004 | 0.006 | 0.010 | 0.032 | 0.037 | 0.040 | 0.061 | 0.048 | 0.055 | 0.063 | 0.101 | 0.059 | 0.066 | 0.074 | 0.111 | 0.139 | 0.162 | 0.165 |
| 112 | INT T OPER RES | 2 | 224 | | 0.481 | 11 | 5 | 0.031 | 0.028 | 0.023 | 0.033 | 0.071 | 0.067 | 0.076 | 0.110 | 0.103 | 0.094 | 0.107 | 0.143 | 0.112 | 0.094 | 0.107 | 0.143 | 0.196 | 0.189 | 0.214 |
| 113 | J GLOBAL OPTIM | 2 | 752 | 1.547 | 1.355 | 21 | 13 | 0.074 | 0.074 | 0.076 | 0.076 | 0.168 | 0.167 | 0.171 | 0.184 | 0.230 | 0.228 | 0.235 | 0.257 | 0.267 | 0.268 | 0.278 | 0.291 | 0.430 | 0.435 | 0.449 |
| 114 | J IND MANAG OPTIM | 2 | 303 | 0.704 | 0.536 | 10 | 6 | 0.026 | 0.025 | 0.027 | 0.041 | 0.063 | 0.071 | 0.088 | 0.116 | 0.099 | 0.109 | 0.132 | 0.182 | 0.119 | 0.134 | 0.154 | 0.207 | 0.221 | 0.244 | 0.286 |
| 115 | J MANUF SYST | 2 | 199 | 1.858 | 1.847 | 11 | 10 | 0.085 | 0.088 | 0.092 | 0.095 | 0.216 | 0.225 | 0.221 | 0.204 | 0.307 | 0.319 | 0.325 | 0.307 | 0.352 | 0.363 | 0.368 | 0.350 | 0.503 | 0.522 | 0.540 |
| 116 | J OPER MANAG | 2 | 208 | 7.718 | 4.478 | 33 | 20 | 0.495 | 0.469 | 0.419 | 0.400 | 0.702 | 0.663 | 0.596 | 0.525 | 0.726 | 0.691 | 0.632 | 0.563 | 0.764 | 0.737 | 0.662 | 0.575 | 0.851 | 0.834 | 0.787 |
| 117 | J OPTIMIZ THEORY APP | 2 | 848 | 1.396 | 1.406 | 21 | 17 | 0.073 | 0.082 | 0.086 | 0.085 | 0.159 | 0.171 | 0.177 | 0.172 | 0.208 | 0.221 | 0.233 | 0.231 | 0.255 | 0.257 | 0.271 | 0.257 | 0.387 | 0.389 | 0.409 |
| 118 | J QUAL TECHNOL | 2 | 149 | 1.681 | 1.271 | 14 | 7 | 0.060 | 0.063 | 0.060 | 0.017 | 0.161 | 0.161 | 0.145 | 0.086 | 0.195 | 0.188 | 0.181 | 0.121 | 0.268 | 0.241 | 0.241 | 0.190 | 0.409 | 0.411 | 0.410 |
| 119 | J SCHEDULING | 2 | 243 | 1.562 | 1.186 | 13 | 9 | 0.053 | 0.034 | 0.044 | 0.045 | 0.144 | 0.128 | 0.151 | 0.162 | 0.189 | 0.177 | 0.208 | 0.216 | 0.235 | 0.222 | 0.258 | 0.252 | 0.370 | 0.365 | 0.396 |
| 120 | J SIMUL | 2 | 131 | | 0.383 | 8 | 4 | 0.038 | 0.037 | 0.013 | 0.019 | 0.046 | 0.047 | 0.026 | 0.038 | 0.046 | 0.047 | 0.026 | 0.038 | 0.061 | 0.065 | 0.038 | 0.038 | 0.153 | 0.131 | 0.128 |
| 121 | J SYST ENG ELECTRON | 2 | 738 | 0.499 | 0.605 | 11 | 7 | 0.009 | 0.011 | 0.008 | 0.004 | 0.027 | 0.031 | 0.032 | 0.025 | 0.039 | 0.048 | 0.047 | 0.042 | 0.049 | 0.061 | 0.066 | 0.055 | 0.099 | 0.127 | 0.137 |
| 122 | J SYST SCI SYST ENG | 2 | 139 | 0.839 | 0.566 | 7 | 4 | 0.000 | 0.000 | 0.000 | 0.000 | 0.036 | 0.045 | 0.049 | 0.057 | 0.043 | 0.054 | 0.049 | 0.057 | 0.058 | 0.063 | 0.061 | 0.057 | 0.137 | 0.143 | 0.146 |
| 123 | J OPER RES SOC | 2 | 976 | 1.272 | 0.911 | 19 | 9 | 0.027 | 0.029 | 0.031 | 0.033 | 0.074 | 0.077 | 0.079 | 0.085 | 0.096 | 0.098 | 0.104 | 0.106 | 0.129 | 0.127 | 0.142 | 0.142 | 0.220 | 0.217 | 0.254 |
| 124 | M&SOM-MANUF SERV OP | 2 | 215 | 2.692 | 1.450 | 18 | 9 | 0.098 | 0.087 | 0.075 | 0.061 | 0.242 | 0.233 | 0.218 | 0.184 | 0.335 | 0.343 | 0.338 | 0.316 | 0.395 | 0.401 | 0.398 | 0.378 | 0.600 | 0.616 | 0.602 |
| 125 | MANAGE SCI | 2 | 732 | 3.458 | 2.524 | 30 | 18 | 0.186 | 0.183 | 0.199 | 0.199 | 0.377 | 0.378 | 0.369 | 0.355 | 0.451 | 0.458 | 0.461 | 0.472 | 0.522 | 0.517 | 0.515 | 0.528 | 0.657 | 0.661 | 0.655 |
| 126 | MATH METHOD OPER RES | 2 | 228 | 0.819 | 0.539 | 9 | 5 | 0.009 | 0.012 | 0.016 | 0.025 | 0.048 | 0.048 | 0.049 | 0.063 | 0.061 | 0.060 | 0.066 | 0.087 | 0.075 | 0.072 | 0.074 | 0.100 | 0.180 | 0.193 | 0.205 |
| 127 | MATH PROGRAM | 2 | 511 | 2.195 | 1.984 | 25 | 17 | 0.153 | 0.153 | 0.157 | 0.152 | 0.294 | 0.292 | 0.292 | 0.266 | 0.346 | 0.353 | 0.354 | 0.333 | 0.391 | 0.389 | 0.398 | 0.380 | 0.526 | 0.528 | 0.542 |
| 128 | MATH OPER RES | 2 | 226 | 1.259 | 0.924 | 13 | 8 | 0.044 | 0.048 | 0.051 | 0.053 | 0.124 | 0.139 | 0.144 | 0.158 | 0.150 | 0.170 | 0.169 | 0.184 | 0.186 | 0.206 | 0.203 | 0.211 | 0.376 | 0.394 | 0.373 |
| 129 | MIL OPER RES | 2 | 85 | 0.337 | 0.312 | 4 | 2 | 0.000 | 0.000 | 0.000 | 0.000 | 0.000 | 0.000 | 0.000 | 0.000 | 0.000 | 0.000 | 0.000 | 0.000 | 0.012 | 0.000 | 0.000 | 0.000 | 0.024 | 0.016 | 0.023 |
| 130 | NAV RES LOG | 2 | 256 | 1.222 | 0.563 | 13 | 7 | 0.043 | 0.035 | 0.033 | 0.053 | 0.105 | 0.094 | 0.086 | 0.074 | 0.125 | 0.114 | 0.113 | 0.117 | 0.148 | 0.139 | 0.139 | 0.138 | 0.285 | 0.257 | 0.265 |
| 131 | NETWORKS | 2 | 304 | 1.245 | 0.739 | 13 | 7 | 0.026 | 0.034 | 0.028 | 0.036 | 0.092 | 0.088 | 0.078 | 0.090 | 0.118 | 0.118 | 0.112 | 0.126 | 0.161 | 0.155 | 0.162 | 0.162 | 0.263 | 0.261 | 0.279 |
| 132 | NETW SPAT ECON | 2 | 142 | 1.640 | 1.803 | 13 | 8 | 0.099 | 0.096 | 0.129 | 0.204 | 0.218 | 0.219 | 0.247 | 0.327 | 0.275 | 0.281 | 0.294 | 0.367 | 0.310 | 0.325 | 0.341 | 0.367 | 0.444 | 0.456 | 0.471 |
| 133 | OMEGA-INT J MANAGE S | 2 | 424 | 3.626 | 3.190 | 30 | 22 | 0.366 | 0.389 | 0.420 | 0.457 | 0.547 | 0.576 | 0.583 | 0.585 | 0.630 | 0.654 | 0.670 | 0.676 | 0.672 | 0.692 | 0.701 | 0.702 | 0.757 | 0.782 | 0.780 |
| 134 | OPER RES | 2 | 570 | 2.498 | 1.500 | 26 | 14 | 0.133 | 0.121 | 0.105 | 0.097 | 0.279 | 0.271 | 0.252 | 0.227 | 0.335 | 0.333 | 0.314 | 0.280 | 0.389 | 0.385 | 0.357 | 0.324 | 0.542 | 0.540 | 0.526 |
| 135 | OPER RES LETT | 2 | 566 | 0.903 | 0.624 | 12 | 8 | 0.012 | 0.013 | 0.014 | 0.016 | 0.051 | 0.049 | 0.054 | 0.054 | 0.072 | 0.070 | 0.080 | 0.078 | 0.094 | 0.087 | 0.094 | 0.089 | 0.208 | 0.212 | 0.222 |
| 136 | OPTIM CONTR APPL MET | 2 | 212 | 1.293 | 1.535 | 13 | 9 | 0.057 | 0.063 | 0.080 | 0.054 | 0.137 | 0.136 | 0.146 | 0.118 | 0.179 | 0.182 | 0.204 | 0.194 | 0.203 | 0.210 | 0.241 | 0.204 | 0.283 | 0.301 | 0.328 |
| 137 | OPTIMIZATION | 2 | 429 | 0.804 | 0.771 | 11 | 9 | 0.028 | 0.033 | 0.043 | 0.044 | 0.070 | 0.074 | 0.086 | 0.087 | 0.100 | 0.104 | 0.118 | 0.126 | 0.114 | 0.121 | 0.139 | 0.148 | 0.217 | 0.234 | 0.243 |
| 138 | OPTIM ENG | 2 | 163 | 1.101 | 0.955 | 11 | 7 | 0.037 | 0.047 | 0.061 | 0.079 | 0.086 | 0.101 | 0.102 | 0.127 | 0.129 | 0.140 | 0.153 | 0.175 | 0.153 | 0.147 | 0.163 | 0.175 | 0.239 | 0.233 | 0.235 |
| 139 | OPTIM LETT | 2 | 457 | 1.201 | 0.990 | 15 | 9 | 0.070 | 0.069 | 0.074 | 0.078 | 0.127 | 0.124 | 0.122 | 0.122 | 0.182 | 0.186 | 0.187 | 0.193 | 0.232 | 0.235 | 0.238 | 0.233 | 0.361 | 0.371 | 0.371 |
| 140 | OPTIM METHOD SOFTW | 2 | 296 | 1.271 | 1.210 | 13 | 8 | 0.047 | 0.041 | 0.050 | 0.069 | 0.122 | 0.107 | 0.127 | 0.138 | 0.159 | 0.144 | 0.171 | 0.192 | 0.179 | 0.160 | 0.193 | 0.200 | 0.314 | 0.309 | 0.359 |
| 141 | PAC J OPTIM | 2 | 214 | 0.848 | 0.798 | 10 | 6 | 0.023 | 0.029 | 0.030 | 0.032 | 0.042 | 0.051 | 0.045 | 0.054 | 0.047 | 0.057 | 0.053 | 0.065 | 0.061 | 0.074 | 0.068 | 0.086 | 0.084 | 0.080 | 0.075 |
| 142 | PROBAB ENG INFORM SC | 2 | 171 | 0.624 | 0.328 | 8 | 4 | 0.000 | 0.000 | 0.000 | 0.000 | 0.018 | 0.007 | 0.010 | 0.014 | 0.023 | 0.015 | 0.010 | 0.014 | 0.029 | 0.022 | 0.010 | 0.014 | 0.064 | 0.051 | 0.029 |
| 143 | P I MECH ENG O-J RIS | 2 | 212 | | 0.775 | 7 | 6 | 0.000 | 0.000 | 0.000 | 0.000 | 0.009 | 0.011 | 0.007 | 0.009 | 0.019 | 0.022 | 0.020 | 0.027 | 0.047 | 0.056 | 0.060 | 0.080 | 0.113 | 0.117 | 0.120 |
| 144 | PROD OPER MANAG | 2 | 352 | 2.378 | 1.759 | 16 | 12 | 0.017 | 0.013 | 0.008 | 0.006 | 0.082 | 0.073 | 0.063 | 0.050 | 0.125 | 0.113 | 0.103 | 0.072 | 0.213 | 0.203 | 0.194 | 0.156 | 0.366 | 0.342 | 0.333 |
| 145 | PROD PLAN CONTROL | 2 | 328 | 1.171 | 0.991 | 14 | 10 | 0.003 | 0.004 | 0.005 | 0.007 | 0.030 | 0.031 | 0.039 | 0.021 | 0.055 | 0.046 | 0.049 | 0.028 | 0.098 | 0.095 | 0.108 | 0.091 | 0.165 | 0.145 | 0.162 |
| 146 | QUAL RELIAB ENG INT | 2 | 456 | 0.997 | 0.994 | 14 | 11 | 0.000 | 0.000 | 0.000 | 0.000 | 0.035 | 0.037 | 0.038 | 0.026 | 0.072 | 0.080 | 0.089 | 0.090 | 0.112 | 0.117 | 0.126 | 0.127 | 0.193 | 0.200 | 0.218 |
| 147 | QUAL TECHNOL QUANT M | 2 | 156 | | 0.339 | 6 | 4 | 0.000 | 0.000 | 0.000 | 0.000 | 0.000 | 0.000 | 0.000 | 0.000 | 0.006 | 0.008 | 0.011 | 0.016 | 0.013 | 0.016 | 0.021 | 0.032 | 0.058 | 0.066 | 0.053 |
| 148 | QUEUEING SYST | 2 | 251 | 0.911 | 0.602 | 10 | 5 | 0.000 | 0.000 | 0.000 | 0.000 | 0.004 | 0.005 | 0.006 | 0.011 | 0.032 | 0.030 | 0.032 | 0.043 | 0.048 | 0.045 | 0.052 | 0.074 | 0.108 | 0.090 | 0.090 |
| 149 | RAIRO-OPER RES | 2 | 109 | 0.364 | 0.333 | 5 | 3 | 0.000 | 0.000 | 0.000 | 0.000 | 0.009 | 0.013 | 0.019 | 0.000 | 0.018 | 0.013 | 0.019 | 0.000 | 0.018 | 0.013 | 0.019 | 0.000 | 0.028 | 0.013 | 0.019 |
| 150 | RELIAB ENG SYST SAFE | 2 | 884 | 2.593 | 2.048 | 29 | 17 | 0.044 | 0.038 | 0.037 | 0.043 | 0.171 | 0.148 | 0.142 | 0.152 | 0.241 | 0.217 | 0.211 | 0.212 | 0.320 | 0.309 | 0.314 | 0.332 | 0.484 | 0.470 | 0.477 |
| 151 | SAFETY SCI | 2 | 999 | 2.020 | 1.672 | 26 | 16 | 0.025 | 0.023 | 0.020 | 0.017 | 0.116 | 0.116 | 0.115 | 0.110 | 0.166 | 0.157 | 0.157 | 0.151 | 0.227 | 0.218 | 0.222 | 0.224 | 0.349 | 0.343 | 0.347 |
| 152 | SORT-STAT OPER RES T | 2 | 65 | 0.807 | 0.962 | 5 | 4 | 0.015 | 0.019 | 0.024 | 0.038 | 0.031 | 0.019 | 0.024 | 0.038 | 0.062 | 0.056 | 0.071 | 0.115 | 0.077 | 0.074 | 0.071 | 0.115 | 0.108 | 0.111 | 0.119 |
| 153 | STUD INFORM CONTROL | 2 | 219 | 0.500 | 0.605 | 8 | 6 | 0.005 | 0.006 | 0.007 | 0.011 | 0.027 | 0.034 | 0.037 | 0.043 | 0.032 | 0.039 | 0.045 | 0.053 | 0.046 | 0.056 | 0.067 | 0.085 | 0.068 | 0.084 | 0.097 |
| 154 | SYST CONTROL LETT | 2 | 684 | 2.345 | 1.886 | 27 | 19 | 0.054 | 0.055 | 0.048 | 0.047 | 0.129 | 0.130 | 0.120 | 0.112 | 0.180 | 0.178 | 0.164 | 0.162 | 0.224 | 0.228 | 0.219 | 0.227 | 0.348 | 0.357 | 0.348 |
| 155 | SYSTEMS ENG | 2 | 152 | 1.072 | 0.923 | 8 | 6 | 0.000 | 0.000 | 0.000 | 0.000 | 0.013 | 0.008 | 0.010 | 0.014 | 0.033 | 0.023 | 0.020 | 0.028 | 0.072 | 0.068 | 0.069 | 0.099 | 0.138 | 0.143 | 0.139 |
| 156 | TECHNOVATION | 2 | 333 | 3.251 | 2.704 | 25 | 14 | 0.060 | 0.044 | 0.038 | 0.000 | 0.225 | 0.196 | 0.141 | 0.092 | 0.291 | 0.248 | 0.179 | 0.126 | 0.414 | 0.376 | 0.337 | 0.294 | 0.565 | 0.528 | 0.467 |
| 157 | TOP | 2 | 197 | 0.955 | 0.766 | 7 | 5 | 0.005 | 0.006 | 0.000 | 0.000 | 0.015 | 0.019 | 0.016 | 0.011 | 0.036 | 0.044 | 0.040 | 0.043 | 0.046 | 0.056 | 0.056 | 0.054 | 0.081 | 0.094 | 0.095 |
| 158 | TRANSPORT RES B-METH | 2 | 495 | 4.439 | 3.894 | 32 | 19 | 0.099 | 0.088 | 0.072 | 0.056 | 0.333 | 0.328 | 0.290 | 0.238 | 0.430 | 0.409 | 0.380 | 0.320 | 0.513 | 0.498 | 0.484 | 0.424 | 0.673 | 0.653 | 0.643 |
| 159 | TRANSPORT RES E-LOG | 2 | 448 | 2.943 | 2.193 | 25 | 18 | 0.063 | 0.060 | 0.057 | 0.047 | 0.210 | 0.202 | 0.175 | 0.183 | 0.277 | 0.262 | 0.225 | 0.236 | 0.337 | 0.322 | 0.289 | 0.288 | 0.502 | 0.484 | 0.418 |
| 160 | TRANSPORT SCI | 2 | 179 | 2.913 | 2.294 | 18 | 11 | 0.112 | 0.120 | 0.150 | 0.208 | 0.251 | 0.246 | 0.271 | 0.333 | 0.324 | 0.324 | 0.336 | 0.375 | 0.419 | 0.430 | 0.430 | 0.486 | 0.587 | 0.592 | 0.598 |
| 161 | ACTA OPHTHALMOL | 3 | 876 | 2.538 | 2.512 | 24 | 16 | 0.036 | 0.034 | 0.034 | 0.035 | 0.086 | 0.084 | 0.086 | 0.082 | 0.104 | 0.100 | 0.098 | 0.091 | 0.156 | 0.150 | 0.147 | 0.137 | 0.208 | 0.197 | 0.191 |
| 162 | AM J OPHTHALMOL | 3 | 1299 | 4.535 | 4.021 | 48 | 26 | 0.173 | 0.168 | 0.177 | 0.196 | 0.312 | 0.309 | 0.323 | 0.347 | 0.357 | 0.351 | 0.359 | 0.377 | 0.432 | 0.436 | 0.454 | 0.468 | 0.496 | 0.496 | 0.503 |
| 163 | ARCH OPHTHALMOL-CHIC | 3 | 727 | 4.481 | 4.488 | 42 | 25 | 0.121 | 0.107 | 0.103 | 0.090 | 0.214 | 0.207 | 0.192 | 0.165 | 0.249 | 0.237 | 0.216 | 0.189 | 0.297 | 0.293 | 0.272 | 0.225 | 0.371 | 0.361 | 0.342 |
| 164 | ARQ BRAS OFTALMOL | 3 | 513 | 0.518 | 0.440 | 8 | 6 | 0.002 | 0.002 | 0.003 | 0.005 | 0.012 | 0.014 | 0.016 | 0.015 | 0.013 | 0.016 | 0.019 | 0.015 | 0.028 | 0.037 | 0.048 | 0.049 | 0.045 | 0.056 | 0.064 |
| 165 | BMC OPHTHALMOL | 3 | 225 | 1.520 | 1.075 | 11 | 8 | 0.013 | 0.014 | 0.005 | 0.007 | 0.072 | 0.073 | 0.069 | 0.061 | 0.089 | 0.082 | 0.074 | 0.068 | 0.183 | 0.173 | 0.169 | 0.155 | 0.226 | 0.209 | 0.206 |
| 166 | BRIT J OPHTHALMOL | 3 | 1455 | 2.967 | 2.809 | 38 | 25 | 0.111 | 0.111 | 0.118 | 0.131 | 0.237 | 0.238 | 0.249 | 0.267 | 0.278 | 0.278 | 0.282 | 0.298 | 0.353 | 0.358 | 0.366 | 0.394 | 0.427 | 0.424 | 0.427 |
| 167 | CAN J OPHTHALMOL | 3 | 315 | 1.568 | 1.299 | 13 | 9 | 0.009 | 0.011 | 0.013 | 0.017 | 0.042 | 0.052 | 0.054 | 0.061 | 0.050 | 0.057 | 0.058 | 0.061 | 0.086 | 0.091 | 0.096 | 0.104 | 0.125 | 0.128 | 0.121 |
| 168 | CLIN EXP OPHTHALMOL | 3 | 478 | 2.158 | 1.953 | 21 | 12 | 0.024 | 0.023 | 0.018 | 0.023 | 0.062 | 0.056 | 0.049 | 0.062 | 0.081 | 0.068 | 0.060 | 0.075 | 0.125 | 0.112 | 0.099 | 0.115 | 0.159 | 0.144 | 0.124 |
| 169 | CLIN EXP OPTOM | 3 | 321 | 1.255 | 1.256 | 15 | 12 | 0.044 | 0.049 | 0.057 | 0.073 | 0.094 | 0.101 | 0.118 | 0.142 | 0.113 | 0.118 | 0.130 | 0.155 | 0.157 | 0.165 | 0.187 | 0.228 | 0.196 | 0.205 | 0.218 |
| 170 | CONTACT LENS ANTERIO | 3 | 232 | | 2.000 | 16 | 9 | 0.058 | 0.063 | 0.051 | 0.054 | 0.148 | 0.161 | 0.138 | 0.155 | 0.180 | 0.192 | 0.168 | 0.171 | 0.273 | 0.286 | 0.260 | 0.302 | 0.334 | 0.341 | 0.316 |
| 171 | CORNEA | 3 | 1454 | 2.239 | 2.360 | 29 | 20 | 0.084 | 0.086 | 0.099 | 0.117 | 0.202 | 0.210 | 0.234 | 0.260 | 0.238 | 0.245 | 0.265 | 0.286 | 0.317 | 0.330 | 0.354 | 0.376 | 0.389 | 0.395 | 0.411 |
| 172 | CURR EYE RES | 3 | 718 | 1.751 | 1.663 | 20 | 13 | 0.048 | 0.055 | 0.053 | 0.059 | 0.155 | 0.164 | 0.162 | 0.184 | 0.197 | 0.204 | 0.193 | 0.205 | 0.285 | 0.299 | 0.306 | 0.329 | 0.369 | 0.379 | 0.390 |
| 173 | CURR OPIN OPHTHALMOL | 3 | 213 | 2.704 | 2.638 | 25 | 17 | 0.140 | 0.142 | 0.146 | 0.145 | 0.287 | 0.300 | 0.305 | 0.283 | 0.351 | 0.362 | 0.354 | 0.313 | 0.442 | 0.464 | 0.467 | 0.452 | 0.553 | 0.563 | 0.565 |
| 174 | CUTAN OCUL TOXICOL | 3 | 246 | 0.987 | 0.920 | 9 | 7 | 0.017 | 0.019 | 0.014 | 0.022 | 0.055 | 0.062 | 0.063 | 0.094 | 0.079 | 0.081 | 0.087 | 0.108 | 0.151 | 0.159 | 0.164 | 0.216 | 0.192 | 0.194 | 0.208 |
| 175 | DOC OPHTHALMOL | 3 | 236 | 1.798 | 1.108 | 15 | 7 | 0.049 | 0.047 | 0.045 | 0.054 | 0.098 | 0.090 | 0.089 | 0.098 | 0.110 | 0.104 | 0.102 | 0.107 | 0.182 | 0.175 | 0.185 | 0.214 | 0.239 | 0.232 | 0.210 |
| 176 | EUR J OPHTHALMOL | 3 | 840 | 1.006 | 1.058 | 16 | 11 | 0.013 | 0.013 | 0.015 | 0.017 | 0.049 | 0.051 | 0.063 | 0.064 | 0.058 | 0.058 | 0.070 | 0.070 | 0.102 | 0.104 | 0.124 | 0.128 | 0.152 | 0.153 | 0.171 |
| 177 | EXP EYE RES | 3 | 957 | 3.087 | 3.017 | 37 | 20 | 0.161 | 0.162 | 0.152 | 0.163 | 0.335 | 0.335 | 0.322 | 0.340 | 0.387 | 0.371 | 0.355 | 0.358 | 0.498 | 0.492 | 0.481 | 0.494 | 0.597 | 0.583 | 0.574 |

| | | | | | | | | | | | | | | | | | | | | | | | | | | | |
|---|---|---|---|---|---|---|---|---|---|---|---|---|---|---|---|---|---|---|---|---|---|---|---|---|---|---|---|
| 178 | EYE | 3 | 1102 | 1.943 | 1.897 | 29 | 19 | 0.053 | 0.055 | 0.065 | 0.068 | 0.127 | 0.127 | 0.150 | 0.155 | 0.154 | 0.146 | 0.162 | 0.164 | 0.211 | 0.212 | 0.222 | 0.224 | 0.283 | 0.281 | 0.280 | 0 |
| 179 | EYE CONTACT LENS | 3 | 295 | 1.920 | 1.679 | 16 | 11 | 0.065 | 0.072 | 0.079 | 0.098 | 0.146 | 0.163 | 0.172 | 0.196 | 0.169 | 0.191 | 0.184 | 0.209 | 0.242 | 0.260 | 0.243 | 0.270 | 0.322 | 0.339 | 0.322 | 0 |
| 180 | GRAEF ARCH CLIN EXP | 3 | 1175 | 2.274 | 2.333 | 28 | 18 | 0.073 | 0.069 | 0.074 | 0.076 | 0.179 | 0.178 | 0.189 | 0.203 | 0.216 | 0.204 | 0.207 | 0.217 | 0.305 | 0.302 | 0.315 | 0.325 | 0.369 | 0.359 | 0.354 | 0 |
| 181 | INDIAN J OPHTHALMOL | 3 | 528 | 0.980 | 0.927 | 42 | 25 | 0.121 | 0.108 | 0.105 | 0.093 | 0.216 | 0.209 | 0.193 | 0.168 | 0.251 | 0.239 | 0.218 | 0.194 | 0.297 | 0.294 | 0.273 | 0.227 | 0.376 | 0.367 | 0.350 | 0 |
| 182 | INT J OPHTHALMOL-CHI | 3 | 592 | 0.302 | 0.500 | 6 | 6 | 0.000 | 0.000 | 0.000 | 0.000 | 0.020 | 0.023 | 0.027 | 0.036 | 0.023 | 0.026 | 0.031 | 0.039 | 0.059 | 0.068 | 0.079 | 0.108 | 0.068 | 0.079 | 0.092 | 0 |
| 183 | INVEST OPHTH VIS SCI | 3 | 4803 | 3.754 | 3.661 | 55 | 41 | 0.216 | 0.215 | 0.214 | 0.205 | 0.417 | 0.417 | 0.417 | 0.405 | 0.480 | 0.473 | 0.461 | 0.447 | 0.595 | 0.594 | 0.589 | 0.572 | 0.684 | 0.678 | 0.669 | 0 |
| 184 | JAMA OPHTHALMOL | 3 | 167 | | | 10 | 10 | 0.129 | 0.129 | 0.129 | 0.129 | 0.253 | 0.253 | 0.253 | 0.253 | 0.253 | 0.253 | 0.253 | 0.253 | 0.346 | 0.346 | 0.346 | 0.346 | 0.346 | 0.346 | 0.346 | 0 |
| 185 | JPN J OPHTHALMOL | 3 | 437 | 1.587 | 1.795 | 16 | 12 | 0.025 | 0.033 | 0.042 | 0.038 | 0.096 | 0.111 | 0.141 | 0.168 | 0.121 | 0.139 | 0.170 | 0.196 | 0.168 | 0.199 | 0.247 | 0.288 | 0.208 | 0.235 | 0.292 | 0 |
| 186 | J FR OPHTALMOL | 3 | 568 | 0.455 | 0.361 | 7 | 5 | 0.000 | 0.000 | 0.000 | 0.000 | 0.003 | 0.003 | 0.004 | 0.005 | 0.004 | 0.005 | 0.004 | 0.005 | 0.019 | 0.021 | 0.021 | 0.024 | 0.029 | 0.034 | 0.032 | 0 |
| 187 | J AAPOS | 3 | 709 | 1.250 | 1.142 | 17 | 10 | 0.011 | 0.012 | 0.010 | 0.014 | 0.054 | 0.053 | 0.057 | 0.066 | 0.076 | 0.076 | 0.078 | 0.086 | 0.130 | 0.137 | 0.145 | 0.157 | 0.192 | 0.196 | 0.196 | 0 |
| 188 | J CATARACT REFR SURG | 3 | 1459 | 2.766 | 2.552 | 38 | 25 | 0.116 | 0.117 | 0.124 | 0.121 | 0.226 | 0.228 | 0.241 | 0.245 | 0.261 | 0.256 | 0.267 | 0.268 | 0.330 | 0.335 | 0.350 | 0.362 | 0.395 | 0.396 | 0.408 | 0 |
| 189 | J EYE MOVEMENT RES | 3 | 87 | 1.427 | 1.056 | 4 | 3 | 0.000 | 0.000 | 0.000 | 0.000 | 0.000 | 0.000 | 0.000 | 0.000 | 0.000 | 0.000 | 0.000 | 0.000 | 0.057 | 0.027 | 0.035 | 0.043 | 0.080 | 0.055 | 0.070 | 0 |
| 190 | J GLAUCOMA | 3 | 585 | 2.297 | 2.427 | 22 | 14 | 0.062 | 0.061 | 0.069 | 0.073 | 0.157 | 0.157 | 0.174 | 0.190 | 0.187 | 0.184 | 0.200 | 0.209 | 0.269 | 0.277 | 0.300 | 0.311 | 0.349 | 0.345 | 0.351 | 0 |
| 191 | J NEURO-OPHTHALMOL | 3 | 259 | 1.504 | 1.807 | 15 | 13 | 0.051 | 0.055 | 0.071 | 0.079 | 0.083 | 0.093 | 0.116 | 0.116 | 0.101 | 0.103 | 0.129 | 0.126 | 0.143 | 0.151 | 0.177 | 0.191 | 0.194 | 0.199 | 0.222 | 0 |
| 192 | J OCUL PHARMACOL TH | 3 | 482 | 1.467 | 1.420 | 16 | 10 | 0.054 | 0.056 | 0.059 | 0.067 | 0.129 | 0.132 | 0.142 | 0.162 | 0.154 | 0.145 | 0.153 | 0.178 | 0.236 | 0.235 | 0.252 | 0.269 | 0.320 | 0.315 | 0.309 | 0 |
| 193 | J OPHTHALMOL | 3 | 172 | 1.852 | 1.935 | 10 | 9 | 0.020 | 0.021 | 0.019 | 0.006 | 0.082 | 0.088 | 0.096 | 0.075 | 0.105 | 0.113 | 0.124 | 0.112 | 0.195 | 0.205 | 0.225 | 0.219 | 0.238 | 0.251 | 0.278 | 0 |
| 194 | J PEDIAT OPHTH STRAB | 3 | 250 | 0.868 | 0.731 | 7 | 5 | 0.005 | 0.004 | 0.005 | 0.007 | 0.014 | 0.014 | 0.015 | 0.022 | 0.022 | 0.025 | 0.025 | 0.037 | 0.046 | 0.058 | 0.065 | 0.090 | 0.073 | 0.090 | 0.104 | 0 |
| 195 | J REFRACT SURG | 3 | 588 | 2.743 | 2.781 | 30 | 20 | 0.136 | 0.133 | 0.144 | 0.160 | 0.274 | 0.276 | 0.299 | 0.332 | 0.310 | 0.308 | 0.328 | 0.362 | 0.391 | 0.403 | 0.428 | 0.456 | 0.454 | 0.464 | 0.475 | 0 |
| 196 | J VISION | 3 | 1433 | 3.075 | 2.727 | 35 | 16 | 0.078 | 0.062 | 0.045 | 0.038 | 0.180 | 0.152 | 0.125 | 0.122 | 0.220 | 0.184 | 0.152 | 0.149 | 0.309 | 0.277 | 0.246 | 0.242 | 0.390 | 0.355 | 0.312 | 0 |
| 197 | KLIN MONATSBL AUGENH | 3 | 522 | 0.526 | 0.665 | 10 | 8 | 0.003 | 0.004 | 0.005 | 0.007 | 0.015 | 0.019 | 0.018 | 0.025 | 0.017 | 0.021 | 0.021 | 0.025 | 0.034 | 0.043 | 0.043 | 0.048 | 0.052 | 0.060 | 0.058 | 0 |
| 198 | MOL VIS | 3 | 1568 | 2.433 | 2.245 | 27 | 17 | 0.079 | 0.078 | 0.069 | 0.076 | 0.218 | 0.227 | 0.223 | 0.220 | 0.284 | 0.279 | 0.260 | 0.252 | 0.407 | 0.408 | 0.397 | 0.410 | 0.520 | 0.518 | 0.499 | 0 |
| 199 | OCUL IMMUNOL INFLAMM | 3 | 335 | 1.381 | 1.440 | 16 | 11 | 0.039 | 0.039 | 0.038 | 0.047 | 0.092 | 0.094 | 0.094 | 0.094 | 0.122 | 0.120 | 0.111 | 0.115 | 0.182 | 0.182 | 0.177 | 0.194 | 0.236 | 0.232 | 0.226 | 0 |
| 200 | OCUL SURF | 3 | 75 | 4.039 | 4.212 | 16 | 12 | 0.127 | 0.137 | 0.137 | 0.149 | 0.280 | 0.290 | 0.284 | 0.254 | 0.306 | 0.315 | 0.316 | 0.284 | 0.350 | 0.363 | 0.368 | 0.328 | 0.376 | 0.379 | 0.389 | 0 |
| 201 | OPHTHAL PHYSL OPT | 3 | 351 | 1.743 | 2.664 | 17 | 14 | 0.086 | 0.092 | 0.118 | 0.129 | 0.187 | 0.207 | 0.274 | 0.288 | 0.213 | 0.228 | 0.295 | 0.307 | 0.299 | 0.325 | 0.392 | 0.411 | 0.380 | 0.402 | 0.464 | 0 |
| 202 | OPHTHAL EPIDEMIOL | 3 | 249 | 1.794 | 1.271 | 15 | 9 | 0.051 | 0.052 | 0.038 | 0.043 | 0.099 | 0.103 | 0.094 | 0.094 | 0.150 | 0.150 | 0.132 | 0.120 | 0.278 | 0.296 | 0.283 | 0.265 | 0.363 | 0.366 | 0.314 | 0 |
| 203 | OPHTHALMIC GENET | 3 | 199 | 1.225 | 1.233 | 11 | 7 | 0.004 | 0.000 | 0.000 | 0.000 | 0.088 | 0.096 | 0.101 | 0.116 | 0.115 | 0.118 | 0.129 | 0.158 | 0.164 | 0.171 | 0.187 | 0.242 | 0.243 | 0.257 | 0.273 | 0 |
| 204 | OPHTHAL PLAST RECONS | 3 | 818 | 0.895 | 0.914 | 12 | 10 | 0.005 | 0.006 | 0.007 | 0.011 | 0.043 | 0.048 | 0.057 | 0.072 | 0.048 | 0.054 | 0.062 | 0.076 | 0.088 | 0.098 | 0.110 | 0.133 | 0.130 | 0.139 | 0.154 | 0 |
| 205 | OPHTHALMIC RES | 3 | 303 | 1.312 | 1.376 | 15 | 11 | 0.042 | 0.042 | 0.033 | 0.050 | 0.110 | 0.117 | 0.100 | 0.108 | 0.140 | 0.136 | 0.114 | 0.129 | 0.200 | 0.201 | 0.190 | 0.209 | 0.275 | 0.280 | 0.271 | 0 |
| 206 | OPHTHAL SURG LAS IM | 3 | 464 | 1.088 | 1.318 | 12 | 8 | 0.012 | 0.008 | 0.007 | 0.005 | 0.059 | 0.031 | 0.036 | 0.045 | 0.059 | 0.062 | 0.064 | 0.070 | 0.107 | 0.119 | 0.132 | 0.139 | 0.168 | 0.182 | 0.196 | 0 |
| 207 | OPHTHALMOLOGE | 3 | 710 | 0.676 | 0.719 | 12 | 8 | 0.002 | 0.003 | 0.002 | 0.003 | 0.028 | 0.027 | 0.017 | 0.017 | 0.039 | 0.037 | 0.025 | 0.025 | 0.060 | 0.056 | 0.045 | 0.047 | 0.095 | 0.094 | 0.081 | 0 |
| 208 | OPHTHALMOLOGICA | 3 | 310 | 1.454 | 1.867 | 16 | 12 | 0.045 | 0.049 | 0.051 | 0.055 | 0.123 | 0.132 | 0.141 | 0.109 | 0.147 | 0.150 | 0.161 | 0.130 | 0.207 | 0.209 | 0.225 | 0.193 | 0.269 | 0.266 | 0.280 | 0 |
| 209 | OPHTHALMOLOGY | 3 | 1687 | 6.195 | 6.170 | 61 | 45 | 0.246 | 0.242 | 0.244 | 0.246 | 0.383 | 0.383 | 0.385 | 0.388 | 0.420 | 0.413 | 0.414 | 0.412 | 0.481 | 0.476 | 0.482 | 0.484 | 0.524 | 0.517 | 0.520 | 0 |
| 210 | OPTOMETRY VISION SCI | 3 | 819 | 2.331 | 2.038 | 25 | 15 | 0.067 | 0.068 | 0.057 | 0.068 | 0.170 | 0.171 | 0.165 | 0.178 | 0.203 | 0.199 | 0.190 | 0.195 | 0.285 | 0.290 | 0.286 | 0.297 | 0.366 | 0.363 | 0.359 | 0 |
| 211 | OPTOMETRY | 3 | 150 | 1.399 | 1.339 | 11 | 6 | 0.012 | 0.010 | 0.019 | 0.083 | 0.023 | 0.024 | 0.038 | 0.083 | 0.023 | 0.024 | 0.038 | 0.083 | 0.031 | 0.031 | 0.038 | 0.083 | 0.045 | 0.045 | 0.056 | 0 |
| 212 | PROG RETIN EYE RES | 3 | 135 | 11.207 | 9.897 | 38 | 24 | 0.701 | 0.701 | 0.713 | 0.707 | 0.841 | 0.843 | 0.851 | 0.853 | 0.879 | 0.881 | 0.881 | 0.867 | 0.904 | 0.910 | 0.911 | 0.907 | 0.924 | 0.925 | 0.921 | 0 |
| 213 | RETINA-J RET VIT DIS | 3 | 1185 | 3.047 | 3.177 | 36 | 26 | 0.124 | 0.131 | 0.130 | 0.140 | 0.237 | 0.245 | 0.250 | 0.261 | 0.271 | 0.275 | 0.271 | 0.279 | 0.351 | 0.368 | 0.367 | 0.381 | 0.424 | 0.440 | 0.439 | 0 |
| 214 | REV BRAS OFTALMOL | 3 | 313 | 0.192 | 0.163 | 4 | 3 | 0.003 | 0.003 | 0.004 | 0.006 | 0.005 | 0.006 | 0.008 | 0.012 | 0.005 | 0.006 | 0.008 | 0.012 | 0.008 | 0.010 | 0.012 | 0.012 | 0.008 | 0.010 | 0.012 | 0 |
| 215 | SEMIN OPHTHALMOL | 3 | 229 | | 1.196 | 15 | 9 | 0.031 | 0.026 | 0.027 | 0.016 | 0.096 | 0.090 | 0.087 | 0.056 | 0.133 | 0.115 | 0.092 | 0.056 | 0.181 | 0.171 | 0.152 | 0.103 | 0.239 | 0.226 | 0.201 | 0 |
| 216 | SURV OPHTHALMOL | 3 | 2 | 3.686 | 3.507 | 27 | 16 | 0.179 | 0.180 | 0.178 | 0.202 | 0.319 | 0.322 | 0.306 | 0.312 | 0.352 | 0.351 | 0.318 | 0.330 | 0.388 | 0.400 | 0.376 | 0.376 | 0.414 | 0.415 | 0.389 | 0 |
| 217 | VISION RES | 3 | 1157 | 2.551 | 2.381 | 36 | 22 | 0.103 | 0.099 | 0.097 | 0.075 | 0.241 | 0.233 | 0.239 | 0.216 | 0.284 | 0.273 | 0.271 | 0.236 | 0.374 | 0.371 | 0.374 | 0.351 | 0.489 | 0.474 | 0.481 | 0 |
| 218 | VISUAL NEUROSCI | 3 | 136 | 1.705 | 1.676 | 14 | 11 | 0.076 | 0.063 | 0.064 | 0.095 | 0.164 | 0.151 | 0.164 | 0.143 | 0.193 | 0.183 | 0.191 | 0.143 | 0.240 | 0.230 | 0.236 | 0.175 | 0.351 | 0.341 | 0.336 | 0 |
| 219 | ADV ENERGY MATER | 4 | 553 | 14.442 | 14.385 | 67 | 67 | 0.635 | 0.635 | 0.635 | 0.624 | 0.821 | 0.821 | 0.821 | 0.825 | 0.866 | 0.866 | 0.866 | 0.863 | 0.913 | 0.913 | 0.913 | 0.912 | 0.948 | 0.948 | 0.948 | 0 |
| 220 | ADV FUNCT MATER | 4 | 2.72 | 11.210 | 10.439 | 111 | 76 | 0.507 | 0.498 | 0.488 | 0.480 | 0.747 | 0.746 | 0.742 | 0.737 | 0.800 | 0.796 | 0.791 | 0.788 | 0.855 | 0.856 | 0.857 | 0.861 | 0.910 | 0.911 | 0.913 | 0 |
| 221 | ADV MATER | 4 | 4096 | 15.581 | 15.409 | 175 | 112 | 0.594 | 0.598 | 0.599 | 0.600 | 0.796 | 0.795 | 0.798 | 0.798 | 0.843 | 0.842 | 0.845 | 0.844 | 0.888 | 0.888 | 0.893 | 0.897 | 0.928 | 0.927 | 0.928 | 0 |
| 222 | ADV COND MATTER PHYS | 4 | 198 | 1.114 | 1.013 | 10 | 6 | 0.005 | 0.005 | 0.006 | 0.007 | 0.035 | 0.036 | 0.031 | 0.037 | 0.051 | 0.052 | 0.038 | 0.045 | 0.091 | 0.093 | 0.069 | 0.075 | 0.141 | 0.145 | 0.126 | 0 |
| 223 | APPL SURF SCI | 4 | 8.66 | 2.469 | 2.538 | 47 | 34 | 0.055 | 0.053 | 0.052 | 0.055 | 0.197 | 0.190 | 0.189 | 0.195 | 0.275 | 0.265 | 0.262 | 0.269 | 0.385 | 0.379 | 0.385 | 0.408 | 0.540 | 0.525 | 0.532 | 0 |
| 224 | CHEM VAPOR DEPOS | 4 | 242 | 1.816 | 1.371 | 15 | 9 | 0.021 | 0.010 | 0.014 | 0.022 | 0.083 | 0.073 | 0.069 | 0.086 | 0.161 | 0.151 | 0.145 | 0.183 | 0.215 | 0.214 | 0.207 | 0.247 | 0.302 | 0.292 | 0.283 | 0 |
| 225 | CONDENS MATTER PHYS | 4 | 310 | 0.714 | 0.771 | 9 | 5 | 0.000 | 0.000 | 0.000 | 0.000 | 0.010 | 0.004 | 0.000 | 0.000 | 0.016 | 0.008 | 0.005 | 0.007 | 0.032 | 0.028 | 0.027 | 0.036 | 0.071 | 0.060 | 0.059 | 0 |
| 226 | EUR PHYS J B | 4 | 2132 | 1.515 | 1.463 | 31 | 17 | 0.020 | 0.013 | 0.012 | 0.010 | 0.070 | 0.055 | 0.055 | 0.045 | 0.108 | 0.087 | 0.085 | 0.074 | 0.165 | 0.149 | 0.153 | 0.153 | 0.261 | 0.241 | 0.246 | 0 |
| 227 | FERROELECTRICS | 4 | 1786 | 0.477 | 0.433 | 13 | 7 | 0.000 | 0.000 | 0.000 | 0.000 | 0.003 | 0.001 | 0.001 | 0.002 | 0.008 | 0.005 | 0.002 | 0.003 | 0.018 | 0.016 | 0.014 | 0.016 | 0.047 | 0.044 | 0.045 | 0 |
| 228 | FERROELECTRICS LETT | 4 | 78 | 0.487 | 0.455 | 6 | 4 | 0.000 | 0.000 | 0.000 | 0.000 | 0.000 | 0.000 | 0.000 | 0.000 | 0.026 | 0.000 | 0.000 | 0.000 | 0.051 | 0.016 | 0.021 | 0.032 | 0.077 | 0.048 | 0.042 | 0 |
| 229 | IEEE T SEMICONDUCT M | 4 | 350 | 1.018 | 0.977 | 12 | 9 | 0.006 | 0.007 | 0.009 | 0.006 | 0.034 | 0.043 | 0.051 | 0.052 | 0.051 | 0.054 | 0.060 | 0.065 | 0.066 | 0.069 | 0.078 | 0.084 | 0.123 | 0.123 | 0.134 | 0 |
| 230 | INTEGR FERROELECTR | 4 | 877 | 0.381 | 0.371 | 9 | 6 | 0.000 | 0.000 | 0.000 | 0.000 | 0.002 | 0.003 | 0.003 | 0.000 | 0.005 | 0.005 | 0.005 | 0.003 | 0.010 | 0.009 | 0.008 | 0.008 | 0.035 | 0.030 | 0.034 | 0 |
| 231 | INT J MOD PHYS B | 4 | 2117 | 0.359 | 0.455 | 15 | 11 | 0.005 | 0.007 | 0.009 | 0.013 | 0.014 | 0.018 | 0.025 | 0.036 | 0.018 | 0.022 | 0.031 | 0.043 | 0.034 | 0.042 | 0.056 | 0.079 | 0.059 | 0.071 | 0.091 | 0 |
| 232 | IONICS | 4 | 694 | 1.736 | 1.836 | 19 | 13 | 0.020 | 0.019 | 0.019 | 0.020 | 0.104 | 0.105 | 0.102 | 0.111 | 0.146 | 0.144 | 0.145 | 0.151 | 0.200 | 0.208 | 0.216 | 0.226 | 0.347 | 0.354 | 0.371 | 0 |
| 233 | J COMPUT THEOR NANOS | 4 | 1782 | 0.855 | 1.032 | 19 | 14 | 0.012 | 0.012 | 0.014 | 0.019 | 0.043 | 0.047 | 0.053 | 0.069 | 0.067 | 0.074 | 0.078 | 0.101 | 0.097 | 0.109 | 0.120 | 0.154 | 0.166 | 0.177 | 0.195 | 0 |
| 234 | J LOW TEMP PHYS | 4 | 1023 | 0.927 | 1.036 | 17 | 12 | 0.004 | 0.003 | 0.002 | 0.002 | 0.052 | 0.050 | 0.053 | 0.060 | 0.079 | 0.075 | 0.078 | 0.091 | 0.117 | 0.108 | 0.117 | 0.140 | 0.197 | 0.188 | 0.209 | 0 |
| 235 | J MAGN | 4 | 325 | | 0.318 | 7 | 4 | 0.000 | 0.000 | 0.000 | 0.000 | 0.003 | 0.004 | 0.004 | 0.007 | 0.003 | 0.004 | 0.004 | 0.007 | 0.009 | 0.007 | 0.008 | 0.014 | 0.034 | 0.028 | 0.033 | 0 |
| 236 | J MAGN MAGN MATER | 4 | 3619 | 1.892 | 2.002 | 40 | 24 | 0.038 | 0.036 | 0.034 | 0.028 | 0.137 | 0.132 | 0.140 | 0.141 | 0.185 | 0.178 | 0.191 | 0.192 | 0.258 | 0.256 | 0.282 | 0.292 | 0.389 | 0.389 | 0.423 | 0 |
| 237 | J MATER SCI-MATER EL | 4 | 1935 | 1.584 | 1.966 | 22 | 17 | 0.021 | 0.023 | 0.023 | 0.028 | 0.089 | 0.097 | 0.100 | 0.108 | 0.135 | 0.146 | 0.145 | 0.155 | 0.219 | 0.236 | 0.243 | 0.264 | 0.348 | 0.367 | 0.374 | 0 |
| 238 | J NANOSCI NANOTECHNO | 4 | 7531 | 1.199 | 1.339 | 34 | 26 | 0.013 | 0.015 | 0.015 | 0.019 | 0.057 | 0.059 | 0.063 | 0.082 | 0.088 | 0.088 | 0.092 | 0.120 | 0.145 | 0.150 | 0.161 | 0.211 | 0.244 | 0.249 | 0.264 | 0 |

| | | | | | | | | | | | | | | | | | | | | | | | | | | |
|---|---|---|---|---|---|---|---|---|---|---|---|---|---|---|---|---|---|---|---|---|---|---|---|---|---|---|
| 239 | J PHYS CHEM SOLIDS | 4 | 1462 | 1.635 | 1.594 | 27 | 17 | 0.023 | 0.022 | 0.021 | 0.026 | 0.128 | 0.126 | 0.125 | 0.141 | 0.177 | 0.171 | 0.159 | 0.183 | 0.246 | 0.244 | 0.242 | 0.284 | 0.378 | 0.369 | 0.372 |
| 240 | J PHYS-CONDENS MAT | 4 | 6014 | 2.335 | 2.223 | 55 | 33 | 0.049 | 0.047 | 0.044 | 0.044 | 0.145 | 0.145 | 0.142 | 0.150 | 0.205 | 0.201 | 0.197 | 0.207 | 0.273 | 0.278 | 0.282 | 0.310 | 0.395 | 0.400 | 0.408 |
| 241 | J SUPERCOND NOV MAGN | 4 | 1841 | 0.854 | 0.930 | 17 | 14 | 0.010 | 0.009 | 0.011 | 0.013 | 0.034 | 0.034 | 0.038 | 0.043 | 0.047 | 0.046 | 0.051 | 0.059 | 0.075 | 0.076 | 0.084 | 0.100 | 0.139 | 0.141 | 0.158 |
| 242 | LASER PHOTONICS REV | 4 | 275 | 9.177 | 9.313 | 40 | 30 | 0.353 | 0.347 | 0.356 | 0.315 | 0.538 | 0.525 | 0.521 | 0.493 | 0.589 | 0.572 | 0.559 | 0.541 | 0.673 | 0.661 | 0.665 | 0.664 | 0.775 | 0.780 | 0.787 |
| 243 | MATER SCI ENG B-ADV | 4 | 1627 | 1.862 | 2.122 | 29 | 21 | 0.026 | 0.030 | 0.034 | 0.027 | 0.120 | 0.126 | 0.145 | 0.139 | 0.186 | 0.186 | 0.204 | 0.185 | 0.263 | 0.272 | 0.294 | 0.282 | 0.392 | 0.401 | 0.423 |
| 244 | MAT SCI SEMICON PROC | 4 | 598 | 1.408 | 1.761 | 13 | 12 | 0.028 | 0.031 | 0.033 | 0.031 | 0.107 | 0.116 | 0.132 | 0.132 | 0.161 | 0.174 | 0.192 | 0.191 | 0.259 | 0.277 | 0.305 | 0.314 | 0.411 | 0.429 | 0.471 |
| 245 | METALLOFIZ NOV TEKH+ | 4 | 781 | 0.098 | 0.109 | 4 | 3 | 0.000 | 0.000 | 0.000 | 0.000 | 0.000 | 0.000 | 0.000 | 0.000 | 0.000 | 0.000 | 0.000 | 0.000 | 0.001 | 0.002 | 0.002 | 0.003 | 0.008 | 0.009 | 0.010 |
| 246 | MOD PHYS LETT B | 4 | 1.53 | 0.485 | 0.687 | 16 | 9 | 0.005 | 0.004 | 0.004 | 0.004 | 0.016 | 0.016 | 0.020 | 0.020 | 0.025 | 0.024 | 0.027 | 0.031 | 0.045 | 0.050 | 0.064 | 0.076 | 0.084 | 0.095 | 0.117 |
| 247 | NANO LETT | 4 | 4775 | 14.452 | 12.940 | 166 | 106 | 0.582 | 0.575 | 0.555 | 0.529 | 0.795 | 0.791 | 0.782 | 0.765 | 0.845 | 0.839 | 0.831 | 0.822 | 0.888 | 0.885 | 0.882 | 0.877 | 0.931 | 0.927 | 0.925 |
| 248 | NAT MATER | 4 | 1482 | 41.775 | 36.425 | 145 | 97 | 0.459 | 0.461 | 0.469 | 0.481 | 0.530 | 0.528 | 0.532 | 0.546 | 0.546 | 0.545 | 0.552 | 0.572 | 0.565 | 0.567 | 0.577 | 0.599 | 0.589 | 0.592 | 0.600 |
| 249 | PHASE TRANSIT | 4 | 518 | 0.955 | 1.044 | 14 | 11 | 0.015 | 0.014 | 0.018 | 0.017 | 0.039 | 0.036 | 0.045 | 0.048 | 0.052 | 0.052 | 0.060 | 0.061 | 0.079 | 0.079 | 0.095 | 0.104 | 0.174 | 0.176 | 0.199 |
| 250 | PHILOS MAG | 4 | 1426 | 1.539 | 1.427 | 25 | 16 | 0.013 | 0.013 | 0.010 | 0.011 | 0.072 | 0.068 | 0.056 | 0.049 | 0.109 | 0.103 | 0.088 | 0.084 | 0.184 | 0.184 | 0.180 | 0.197 | 0.300 | 0.295 | 0.291 |
| 251 | PHIL MAG LETT | 4 | 449 | 1.137 | 1.268 | 17 | 10 | 0.020 | 0.022 | 0.027 | 0.017 | 0.062 | 0.059 | 0.053 | 0.035 | 0.076 | 0.076 | 0.072 | 0.064 | 0.138 | 0.134 | 0.129 | 0.145 | 0.258 | 0.258 | 0.239 |
| 252 | PHYSICA B | 4 | 4674 | 1.133 | 1.276 | 30 | 20 | 0.012 | 0.014 | 0.013 | 0.013 | 0.062 | 0.070 | 0.070 | 0.069 | 0.091 | 0.099 | 0.098 | 0.097 | 0.144 | 0.161 | 0.165 | 0.170 | 0.239 | 0.264 | 0.268 |
| 253 | PHYSICA E | 4 | 1951 | 1.436 | 1.856 | 35 | 22 | 0.037 | 0.038 | 0.037 | 0.034 | 0.108 | 0.105 | 0.118 | 0.124 | 0.153 | 0.145 | 0.162 | 0.171 | 0.206 | 0.202 | 0.237 | 0.262 | 0.313 | 0.307 | 0.364 |
| 254 | PHYS STATUS SOLIDI A | 4 | 2295 | 1.476 | 1.525 | 31 | 19 | 0.018 | 0.018 | 0.014 | 0.012 | 0.072 | 0.070 | 0.074 | 0.081 | 0.105 | 0.104 | 0.106 | 0.117 | 0.171 | 0.176 | 0.184 | 0.214 | 0.291 | 0.295 | 0.309 |
| 255 | PHYS STATUS SOLIDI B | 4 | 2.24 | 1.405 | 1.605 | 32 | 22 | 0.021 | 0.023 | 0.023 | 0.014 | 0.082 | 0.086 | 0.095 | 0.092 | 0.110 | 0.113 | 0.122 | 0.115 | 0.155 | 0.163 | 0.183 | 0.191 | 0.250 | 0.263 | 0.289 |
| 256 | PHYS STATUS SOLIDI-R | 4 | 776 | 2.187 | 2.343 | 27 | 19 | 0.058 | 0.056 | 0.058 | 0.057 | 0.161 | 0.167 | 0.179 | 0.173 | 0.213 | 0.218 | 0.227 | 0.221 | 0.295 | 0.306 | 0.318 | 0.314 | 0.442 | 0.452 | 0.473 |
| 257 | PHYS REV B | 4 | 29084 | 3.564 | 3.664 | 120 | 79 | 0.125 | 0.123 | 0.121 | 0.123 | 0.309 | 0.309 | 0.309 | 0.318 | 0.386 | 0.381 | 0.378 | 0.389 | 0.476 | 0.479 | 0.489 | 0.513 | 0.614 | 0.614 | 0.624 |
| 258 | PHYS CHEM LIQ | 4 | 342 | 0.530 | 0.517 | 9 | 8 | 0.003 | 0.004 | 0.005 | 0.008 | 0.018 | 0.022 | 0.029 | 0.038 | 0.035 | 0.044 | 0.049 | 0.061 | 0.061 | 0.074 | 0.078 | 0.084 | 0.105 | 0.129 | 0.137 |
| 259 | PHYS SOLID STATE+ | 4 | 2031 | 0.699 | 0.782 | 14 | 11 | 0.002 | 0.002 | 0.002 | 0.003 | 0.010 | 0.011 | 0.012 | 0.015 | 0.018 | 0.020 | 0.022 | 0.024 | 0.038 | 0.044 | 0.048 | 0.059 | 0.111 | 0.121 | 0.139 |
| 260 | PLASMA PROCESS POLYM | 4 | 726 | 2.799 | 2.964 | 29 | 18 | 0.055 | 0.077 | 0.053 | 0.049 | 0.174 | 0.231 | 0.193 | 0.186 | 0.223 | 0.283 | 0.240 | 0.217 | 0.306 | 0.403 | 0.365 | 0.367 | 0.434 | 0.545 | 0.500 |
| 261 | PROG SURF SCI | 4 | 61 | 8.288 | 4.913 | 21 | 11 | 0.328 | 0.255 | 0.216 | 0.160 | 0.541 | 0.489 | 0.432 | 0.400 | 0.623 | 0.596 | 0.541 | 0.520 | 0.738 | 0.745 | 0.703 | 0.720 | 0.820 | 0.809 | 0.757 |
| 262 | RADIAT EFF DEFECT S | 4 | 572 | 0.472 | 0.603 | 9 | 7 | 0.000 | 0.000 | 0.000 | 0.000 | 0.014 | 0.017 | 0.020 | 0.025 | 0.014 | 0.017 | 0.020 | 0.025 | 0.026 | 0.032 | 0.040 | 0.051 | 0.068 | 0.075 | 0.088 |
| 263 | SEMICOND SCI TECH | 4 | 1291 | 1.812 | 2.206 | 31 | 21 | 0.040 | 0.046 | 0.040 | 0.042 | 0.105 | 0.119 | 0.107 | 0.108 | 0.140 | 0.159 | 0.153 | 0.162 | 0.199 | 0.221 | 0.223 | 0.249 | 0.308 | 0.332 | 0.340 |
| 264 | SEMICONDUCTORS+ | 4 | 1.52 | 0.583 | 0.705 | 12 | 9 | 0.003 | 0.002 | 0.002 | 0.003 | 0.010 | 0.011 | 0.015 | 0.019 | 0.019 | 0.021 | 0.025 | 0.031 | 0.036 | 0.042 | 0.053 | 0.072 | 0.087 | 0.097 | 0.117 |
| 265 | SEMICONDUCT SEMIMET | 4 | 52 | 1.400 | 1.742 | 7 | 7 | 0.019 | 0.019 | 0.020 | 0.000 | 0.038 | 0.038 | 0.041 | 0.029 | 0.038 | 0.038 | 0.041 | 0.029 | 0.154 | 0.154 | 0.143 | 0.147 | 0.308 | 0.308 | 0.306 |
| 266 | SMALL | 4 | 2179 | 8.416 | 7.514 | 93 | 53 | 0.353 | 0.335 | 0.324 | 0.326 | 0.606 | 0.600 | 0.596 | 0.601 | 0.677 | 0.666 | 0.661 | 0.661 | 0.763 | 0.761 | 0.767 | 0.783 | 0.846 | 0.843 | 0.845 |
| 267 | SOLID STATE COMMUN | 4 | 2353 | 1.781 | 1.698 | 36 | 23 | 0.035 | 0.031 | 0.037 | 0.044 | 0.112 | 0.107 | 0.120 | 0.107 | 0.161 | 0.153 | 0.167 | 0.150 | 0.213 | 0.210 | 0.230 | 0.215 | 0.329 | 0.325 | 0.354 |
| 268 | SOLID STATE ELECTRON | 4 | 1336 | 1.508 | 1.514 | 26 | 17 | 0.025 | 0.025 | 0.026 | 0.027 | 0.085 | 0.086 | 0.094 | 0.096 | 0.126 | 0.128 | 0.133 | 0.145 | 0.186 | 0.194 | 0.203 | 0.222 | 0.316 | 0.319 | 0.334 |
| 269 | SOLID STATE IONICS | 4 | 1659 | 2.397 | 2.112 | 35 | 23 | 0.052 | 0.048 | 0.044 | 0.043 | 0.183 | 0.176 | 0.165 | 0.152 | 0.250 | 0.238 | 0.223 | 0.210 | 0.352 | 0.350 | 0.340 | 0.350 | 0.503 | 0.499 | 0.494 |
| 270 | SOLID STATE NUCL MAG | 4 | 206 | 2.257 | 2.864 | 19 | 12 | 0.068 | 0.064 | 0.068 | 0.075 | 0.199 | 0.191 | 0.188 | 0.200 | 0.228 | 0.213 | 0.214 | 0.237 | 0.272 | 0.277 | 0.282 | 0.325 | 0.408 | 0.411 | 0.419 |
| 271 | SOLID STATE SCI | 4 | 1604 | 1.883 | 1.679 | 29 | 18 | 0.027 | 0.021 | 0.027 | 0.031 | 0.120 | 0.106 | 0.095 | 0.105 | 0.175 | 0.158 | 0.138 | 0.156 | 0.256 | 0.248 | 0.238 | 0.267 | 0.392 | 0.385 | 0.375 |
| 272 | SOLID STATE TECHNOL | 4 | 541 | 0.230 | 0.247 | 6 | 2 | 0.000 | 0.000 | 0.000 | 0.000 | 0.000 | 0.000 | 0.000 | 0.000 | 0.000 | 0.000 | 0.000 | 0.000 | 0.002 | 0.002 | 0.003 | 0.005 | 0.007 | 0.007 | 0.010 |
| 273 | SUPERCOND SCI TECH | 4 | 1368 | 2.455 | 2.796 | 35 | 21 | 0.056 | 0.054 | 0.047 | 0.051 | 0.178 | 0.181 | 0.168 | 0.177 | 0.235 | 0.237 | 0.225 | 0.234 | 0.314 | 0.322 | 0.321 | 0.340 | 0.450 | 0.460 | 0.452 |
| 274 | SUPERLATTICE MICROST | 4 | 1103 | 1.716 | 1.979 | 26 | 19 | 0.047 | 0.053 | 0.051 | 0.053 | 0.143 | 0.164 | 0.163 | 0.172 | 0.176 | 0.196 | 0.198 | 0.207 | 0.243 | 0.272 | 0.282 | 0.300 | 0.360 | 0.394 | 0.413 |
| 275 | SURF REV LETT | 4 | 393 | 0.398 | 0.367 | 9 | 4 | 0.000 | 0.000 | 0.000 | 0.000 | 0.003 | 0.000 | 0.000 | 0.000 | 0.008 | 0.004 | 0.000 | 0.000 | 0.015 | 0.011 | 0.006 | 0.000 | 0.041 | 0.038 | 0.039 |
| 276 | SURF SCI | 4 | 1794 | 1.876 | 1.870 | 33 | 21 | 0.037 | 0.036 | 0.041 | 0.035 | 0.117 | 0.104 | 0.114 | 0.130 | 0.175 | 0.157 | 0.168 | 0.174 | 0.235 | 0.228 | 0.245 | 0.273 | 0.371 | 0.362 | 0.386 |
| 277 | SYNTHETIC MET | 4 | 2.09 | 2.256 | 2.222 | 33 | 21 | 0.028 | 0.027 | 0.023 | 0.023 | 0.148 | 0.152 | 0.141 | 0.133 | 0.214 | 0.211 | 0.197 | 0.184 | 0.298 | 0.310 | 0.309 | 0.316 | 0.452 | 0.462 | 0.467 |
| 278 | THIN SOLID FILMS | 4 | 7447 | 2.038 | 1.867 | 47 | 28 | 0.026 | 0.021 | 0.018 | 0.018 | 0.109 | 0.100 | 0.095 | 0.093 | 0.158 | 0.142 | 0.135 | 0.136 | 0.227 | 0.215 | 0.219 | 0.228 | 0.365 | 0.352 | 0.362 |

Source: Thomson Reuters Web of Science. WOS Categories: 1 = INFORMATION SCIENCE & LIBRARY SCIENCE; 2 = OPERATIONS RESEARCH & MANAGEMENT SCIENCE; 3 = OPHTHALMOLOGY; 4 = PHYSICS, CONDENSED MATTER.

**SUPPLEMENTARY MATERIAL 2:** Ranking position within each journal category for some impact indicators

| *id* | *Abbreviated Journal Title* | *WOS Category* | *Ranking position* | | | | | |
|---|---|---|---|---|---|---|---|---|
| | | | *JIF5* | *JIF2* | *h5* | *h3* | *pArt_20_2* | *pArt_40_5* |
| 1 | AFR J LIBR ARCH INFO | 1 | 71 | 77 | 76 | 79 | 57 | 81 |
| 2 | ASLIB PROC | 1 | 42 | 47 | 36 | 42 | 57 | 39 |
| 3 | AUST ACAD RES LIBR | 1 | 47 | 50 | 51 | 59 | 57 | 63 |
| 4 | AUST LIBR J | 1 | | 76 | 84 | 79 | 57 | 75 |
| 5 | CAN J INFORM LIB SCI | 1 | 70 | 80 | 73 | 72 | 57 | 70 |
| 6 | COLL RES LIBR | 1 | 32 | 25 | 36 | 28 | 32 | 49 |
| 7 | DATA BASE ADV INF SY | 1 | | 82 | 51 | 59 | 36 | 50 |
| 8 | ECONTENT | 1 | 73 | 84 | 76 | 79 | 57 | 81 |
| 9 | ELECTRON LIBR | 1 | 55 | 75 | 32 | 34 | 57 | 58 |
| 10 | ETHICS INF TECHNOL | 1 | | 51 | 42 | 49 | 50 | 34 |
| 11 | EUR J INFORM SYST | 1 | 13 | 18 | 16 | 14 | 9 | 15 |
| 12 | GOV INFORM Q | 1 | 20 | 11 | 12 | 6 | 15 | 27 |
| 13 | HEALTH INFO LIBR J | 1 | 33 | 37 | 36 | 34 | 29 | 32 |
| 14 | INFORM SOC-ESTUD | 1 | 74 | 79 | 76 | 79 | 57 | 79 |
| 15 | INFORM ORGAN-UK | 1 | 15 | 6 | 30 | 20 | 6 | 13 |
| 16 | INFORM CULT | 1 | 62 | 66 | 76 | 72 | 44 | 48 |
| 17 | INFORM DEV | 1 | 56 | 57 | 51 | 49 | 52 | 65 |
| 18 | INFORM MANAGE-AMSTER | 1 | 7 | 16 | 6 | 11 | 10 | 7 |
| 19 | INFORM PROCESS MANAG | 1 | 27 | 30 | 21 | 20 | 21 | 19 |
| 20 | INFORM RES | 1 | 37 | 43 | 73 | 72 | 57 | 75 |
| 21 | INFORM SOC | 1 | 31 | 35 | 32 | 34 | 42 | 37 |
| 22 | INFORM SYST J | 1 | 12 | 24 | 21 | 28 | 13 | 12 |
| 23 | INFORM SYST RES | 1 | 4 | 7 | 6 | 10 | 8 | 4 |
| 24 | INFORM TECHNOL DEV | 1 | | 58 | 45 | 59 | 57 | 41 |
| 25 | INFORM TECHNOL MANAG | 1 | 36 | 38 | 30 | 11 | 5 | 31 |
| 26 | INFORM TECHNOL PEOPL | 1 | | 36 | 45 | 49 | 57 | 28 |
| 27 | INTERLEND DOC SUPPLY | 1 | 66 | 64 | 61 | 49 | 50 | 38 |
| 28 | INT J COMP-SUPP COLL | 1 | 14 | 15 | 16 | 14 | 11 | 6 |
| 29 | INT J GEOGR INF SCI | 1 | 21 | 20 | 11 | 7 | 16 | 17 |
| 30 | INT J INFORM MANAGE | 1 | 19 | 10 | 9 | 7 | 18 | 22 |
| 31 | INVESTIG BIBLIOTECOL | 1 | 69 | 81 | 76 | 79 | 57 | 74 |
| 32 | J ACAD LIBR | 1 | 40 | 48 | 36 | 34 | 57 | 53 |
| 33 | J COMPUT-MEDIAT COMM | 1 | 3 | 12 | 9 | 14 | 6 | 5 |
| 34 | J DOC | 1 | 29 | 32 | 28 | 24 | 25 | 33 |
| 35 | J GLOB INF MANAG | 1 | 49 | 55 | 51 | 59 | 57 | 44 |
| 36 | J GLOB INF TECH MAN | 1 | | 53 | 68 | 59 | 57 | 60 |
| 37 | J HEALTH COMMUN | 1 | 17 | 14 | 7 | 6 | 17 | 11 |
| 38 | J INF SCI | 1 | 30 | 29 | 16 | 24 | 30 | 20 |
| 39 | J INF TECHNOL | 1 | 2 | 3 | 21 | 34 | 26 | 29 |
| 40 | J INFORMETR | 1 | 6 | 4 | 4 | 5 | 3 | 2 |
| 41 | J KNOWL MANAG | 1 | | 26 | 16 | 14 | 19 | 9 |
| 42 | J LIBR INF SCI | 1 | 43 | 69 | 45 | 59 | 55 | 63 |
| 43 | J MANAGE INFORM SYST | 1 | 8 | 13 | 13 | 14 | 23 | 16 |
| 44 | J ORGAN END USER COM | 1 | | 59 | 61 | 59 | 36 | 45 |
| 45 | J SCHOLARLY PUBL | 1 | 59 | 73 | 68 | 59 | 57 | 69 |
| 46 | J STRATEGIC INF SYST | 1 | 9 | 5 | 16 | 11 | 12 | 10 |
| 47 | J AM MED INFORM ASSN | 1 | 5 | 2 | 2 | 1 | 2 | 3 |
| 48 | J AM SOC INF SCI TEC | 1 | 16 | 9 | 3 | 3 | 14 | 21 |
| 49 | J ASSOC INF SYST | 1 | 11 | 27 | 14 | 28 | 20 | 14 |
| 50 | J MED LIBR ASSOC | 1 | 34 | 34 | 27 | 28 | 44 | 43 |
| 51 | KNOWL MAN RES PRACT | 1 | 35 | 42 | 42 | 28 | 33 | 36 |
| 52 | KNOWL ORGAN | 1 | 52 | 56 | 61 | 49 | 54 | 55 |
| 53 | LEARN PUBL | 1 | 38 | 31 | 32 | 34 | 38 | 42 |
| 54 | LIBR INFORM SC | 1 | 67 | 67 | 76 | 79 | 57 | 73 |
| 55 | LIBR COLLECT ACQUIS | 1 | 51 | 68 | 61 | 59 | 57 | 68 |
| 56 | LIBR HI TECH | 1 | 48 | 61 | 36 | 34 | 48 | 47 |
| 57 | LIBR INFORM SCI RES | 1 | 24 | 22 | 21 | 24 | 34 | 24 |
| 58 | LIBR J | 1 | 64 | 74 | 61 | 49 | 57 | 80 |
| 59 | LIBR QUART | 1 | 41 | 39 | 51 | 59 | 44 | 52 |
| 60 | LIBR RESOUR TECH SER | 1 | 50 | 45 | 51 | 49 | 57 | 57 |
| 61 | LIBR TRENDS | 1 | 54 | 72 | 51 | 49 | 47 | 61 |
| 62 | LIBRI | 1 | 58 | 71 | 68 | 59 | 57 | 66 |
| 63 | MALAYS J LIBR INF SC | 1 | 57 | 65 | 61 | 49 | 57 | 53 |
| 64 | MIS Q EXEC | 1 | 22 | 33 | 1 | 2 | 1 | 1 |
| 65 | MIS QUART | 1 | 1 | 1 | 32 | 42 | 42 | 30 |
| 66 | ONLINE INFORM REV | 1 | 26 | 21 | 21 | 20 | 40 | 40 |
| 67 | PORTAL-LIBR ACAD | 1 | 44 | 44 | 45 | 42 | 52 | 51 |
| 68 | PROF INFORM | 1 | 63 | 60 | 42 | 34 | 49 | 61 |
| 69 | PROGRAM-ELECTRON LIB | 1 | 53 | 52 | 51 | 42 | 35 | 56 |
| 70 | REF USER SERV Q | 1 | 61 | 70 | 61 | 49 | 57 | 72 |
| 71 | RES EVALUAT | 1 | 28 | 23 | 28 | 24 | 28 | 26 |
| 72 | RESTAURATOR | 1 | 60 | 54 | 68 | 59 | 39 | 67 |
| 73 | REV ESP DOC CIENT | 1 | 39 | 40 | 45 | 42 | 57 | 46 |
| 74 | SCIENTIST | 1 | 68 | 63 | 68 | 72 | 57 | 78 |
| 75 | SCIENTOMETRICS | 1 | 18 | 8 | 4 | 3 | 4 | 8 |
| 76 | SERIALS REV | 1 | 45 | 49 | 51 | 42 | 40 | 59 |
| 77 | SOC SCI COMPUT REV | 1 | 23 | 19 | 21 | 20 | 27 | 18 |
| 78 | SOC SCI INFORM | 1 | 46 | 46 | 51 | 42 | 30 | 35 |
| 79 | TELECOMMUN POLICY | 1 | 25 | 28 | 14 | 14 | 24 | 25 |
| 80 | TELEMAT INFORM | 1 | | 41 | 45 | 28 | 22 | 23 |
| 81 | TRANSINFORMACAO | 1 | 72 | 78 | 76 | 72 | 57 | 71 |
| 82 | Z BIBL BIBL | 1 | 75 | 83 | 76 | 72 | 56 | 77 |
| 83 | 4OR-Q J OPER RES | 2 | 43 | 47 | 51 | 59 | 65 | 52 |
| 84 | ANN OPER RES | 2 | 35 | 35 | 21 | 24 | 27 | 37 |
| 85 | APPL STOCH MODEL BUS | 2 | 61 | 66 | 51 | 51 | 52 | 60 |
| 86 | ASIA PAC J OPER RES | 2 | 71 | 78 | 65 | 59 | 65 | 65 |
| 87 | CENT EUR J OPER RES | 2 | 56 | 50 | 58 | 51 | 46 | 45 |
| 88 | COMPUT OPTIM APPL | 2 | 34 | 42 | 25 | 24 | 28 | 25 |
| 89 | COMPUT OPER RES | 2 | 15 | 19 | 3 | 4 | 9 | 5 |
| 90 | CONCURRENT ENG-RES A | 2 | 64 | 67 | 65 | 59 | 51 | 48 |
| 91 | DECIS SUPPORT SYST | 2 | 9 | 9 | 8 | 6 | 13 | 8 |
| 92 | DISCRETE EVENT DYN S | 2 | 49 | 55 | 65 | 59 | 49 | 44 |
| 93 | DISCRETE OPTIM | 2 | 52 | 57 | 51 | 31 | 54 | 46 |
| 94 | ENG ECON | 2 | | 56 | 75 | 69 | 33 | 67 |
| 95 | ENG OPTIMIZ | 2 | 33 | 31 | 31 | 24 | 34 | 31 |
| 96 | EUR J IND ENG | 2 | 30 | 24 | 45 | 31 | 48 | 35 |
| 97 | EUR J OPER RES | 2 | 10 | 15 | 2 | 2 | 8 | 9 |
| 98 | EXPERT SYST APPL | 2 | 16 | 11 | 1 | 1 | 3 | 10 |
| 99 | FLEX SERV MANUF J | 2 | 44 | 26 | 65 | 51 | 17 | 42 |
| 100 | FUZZY OPTIM DECIS MA | 2 | 18 | 38 | 43 | 38 | 7 | 12 |
| 101 | IEEE SYST J | 2 | 21 | 18 | 25 | 24 | 10 | 22 |
| 102 | IIE TRANS | 2 | 28 | 36 | 25 | 30 | 18 | 24 |
| 103 | IMA J MANAG MATH | 2 | 63 | 70 | 65 | 69 | 54 | 57 |
| 104 | INFOR | 2 | 70 | 72 | 73 | 69 | 65 | 75 |
| 105 | INFORMS J COMPUT | 2 | 22 | 34 | 25 | 38 | 23 | 27 |
| 106 | INTERFACES | 2 | 65 | 71 | 51 | 59 | 62 | 64 |
| 107 | INT J COMPUT INTEG M | 2 | 46 | 37 | 35 | 30 | 47 | 38 |
| 108 | INT J INF TECH DECIS | 2 | 25 | 12 | 24 | 20 | 25 | 36 |
| 109 | INT J PROD RES | 2 | 23 | 29 | 16 | 11 | 16 | 19 |
| 110 | INT J SYST SCI | 2 | 24 | 21 | 19 | 16 | 11 | 17 |
| 111 | INT J TECHNOL MANAGE | 2 | 66 | 68 | 58 | 59 | 60 | 61 |
| 112 | INT T OPER RES | 2 | | 69 | 45 | 59 | 43 | 54 |
| 113 | J GLOBAL OPTIM | 2 | 31 | 28 | 17 | 29 | 22 | 21 |
| 114 | J IND MANAG OPTIM | 2 | 62 | 65 | 51 | 51 | 40 | 49 |
| 115 | J MANUF SYST | 2 | 20 | 14 | 45 | 24 | 15 | 15 |
| 116 | J OPER MANAG | 2 | 1 | 1 | 4 | 5 | 2 | 1 |
| 117 | J OPTIMIZ THEORY APP | 2 | 32 | 27 | 17 | 11 | 19 | 26 |
| 118 | J QUAL TECHNOL | 2 | 26 | 30 | 31 | 31 | 56 | 23 |
| 119 | J SCHEDULING | 2 | 29 | 33 | 35 | 30 | 37 | 29 |
| 120 | J SIMUL | 2 | | 73 | 61 | 69 | 52 | 59 |
| 121 | J SYST ENG ELECTRON | 2 | 69 | 60 | 45 | 31 | 64 | 70 |
| 122 | J SYST SCI SYST ENG | 2 | 57 | 62 | 65 | 69 | 65 | 63 |
| 123 | J OPER RES SOC | 2 | 37 | 48 | 19 | 30 | 43 | 50 |
| 124 | M&SOM-MANUF SERV OP | 2 | 8 | 25 | 21 | 30 | 26 | 6 |
| 125 | MANAGE SCI | 2 | 4 | 5 | 6 | 9 | 6 | 4 |
| 126 | MATH METHOD OPER RES | 2 | 58 | 64 | 58 | 59 | 39 | 56 |
| 127 | MATH PROGRAM | 2 | 17 | 10 | 13 | 11 | 12 | 14 |
| 128 | MATH OPER RES | 2 | 39 | 45 | 35 | 38 | 31 | 28 |
| 129 | MIL OPER RES | 2 | 73 | 77 | 78 | 78 | 65 | 78 |
| 130 | NAV RES LOG | 2 | 41 | 63 | 35 | 31 | 31 | 40 |
| 131 | NETWORKS | 2 | 40 | 54 | 35 | 31 | 42 | 43 |
| 132 | NETW SPAT ECON | 2 | 27 | 16 | 35 | 38 | 5 | 20 |
| 133 | OMEGA-INT J MANAGE S | 2 | 3 | 3 | 6 | 3 | 1 | 2 |
| 134 | OPER RES | 2 | 12 | 23 | 11 | 17 | 14 | 13 |
| 135 | OPER RES LETT | 2 | 54 | 58 | 43 | 38 | 58 | 53 |
| 136 | OPTIM CONTR APPL MET | 2 | 36 | 22 | 35 | 30 | 30 | 41 |
| 137 | OPTIMIZATION | 2 | 60 | 52 | 45 | 30 | 38 | 51 |
| 138 | OPTIM ENG | 2 | 47 | 44 | 45 | 31 | 20 | 47 |
| 139 | OPTIM LETT | 2 | 42 | 41 | 30 | 30 | 21 | 32 |
| 140 | OPTIM METHOD SOFTW | 2 | 38 | 32 | 35 | 38 | 24 | 39 |
| 141 | PAC J OPTIM | 2 | 55 | 49 | 51 | 51 | 45 | 71 |
| 142 | PROBAB ENG INFORM SC | 2 | 67 | 76 | 61 | 69 | 65 | 74 |
| 143 | P I MECH ENG O-J RIS | 2 | | 51 | 65 | 51 | 65 | 66 |
| 144 | PROD OPER MANAG | 2 | 13 | 17 | 25 | 20 | 62 | 30 |
| 145 | PROD PLAN CONTROL | 2 | 45 | 40 | 31 | 24 | 61 | 58 |
| 146 | QUAL RELIAB ENG INT | 2 | 50 | 39 | 31 | 22 | 65 | 55 |
| 147 | QUAL TECHNOL QUANT M | 2 | | 74 | 73 | 69 | 65 | 76 |
| 148 | QUEUEING SYST | 2 | 53 | 61 | 51 | 59 | 65 | 68 |
| 149 | RAIRO-OPER RES | 2 | 72 | 75 | 75 | 77 | 65 | 77 |
| 150 | RELIAB ENG SYST SAFE | 2 | 11 | 8 | 8 | 11 | 39 | 18 |
| 151 | SAFETY SCI | 2 | 19 | 20 | 11 | 15 | 56 | 33 |
| 152 | SORT-STAT OPER RES T | 2 | 59 | 43 | 75 | 69 | 41 | 68 |
| 153 | STUD INFORM CONTROL | 2 | 68 | 59 | 61 | 51 | 59 | 73 |
| 154 | SYST CONTROL LETT | 2 | 14 | 13 | 10 | 6 | 35 | 34 |
| 155 | SYSTEMS ENG | 2 | 48 | 46 | 61 | 51 | 65 | 62 |
| 156 | TECHNOVATION | 2 | 5 | 4 | 13 | 17 | 65 | 11 |
| 157 | TOP | 2 | 51 | 53 | 65 | 59 | 65 | 72 |
| 158 | TRANSPORT RES B-METH | 2 | 2 | 2 | 5 | 6 | 29 | 3 |
| 159 | TRANSPORT RES E-LOG | 2 | 6 | 7 | 13 | 9 | 35 | 16 |
| 160 | TRANSPORT SCI | 2 | 7 | 6 | 21 | 22 | 4 | 7 |
| 161 | ACTA OPHTHALMOL | 3 | 16 | 16 | 21 | 18 | 39 | 39 |
| 162 | AM J OPHTHALMOL | 3 | 3 | 5 | 3 | 3 | 5 | 7 |
| 163 | ARCH OPHTHALMOL-CHIC | 3 | 4 | 3 | 4 | 5 | 19 | 19 |
| 164 | ARQ BRAS OFTALMOL | 3 | 52 | 55 | 53 | 52 | 52 | 55 |
| 165 | BMC OPHTHALMOL | 3 | 33 | 45 | 46 | 45 | 47 | 38 |
| 166 | BRIT J OPHTHALMOL | 3 | 11 | 10 | 6 | 5 | 11 | 10 |
| 167 | CAN J OPHTHALMOL | 3 | 32 | 38 | 42 | 40 | 42 | 49 |
| 168 | CLIN EXP OPHTHALMOL | 3 | 22 | 24 | 23 | 26 | 40 | 46 |
| 169 | CLIN EXP OPTOM | 3 | 41 | 40 | 35 | 26 | 25 | 41 |
| 170 | CONTACT LENS ANTERIO | 3 | | 23 | 27 | 40 | 32 | 27 |
| 171 | CORNEA | 3 | 21 | 19 | 14 | 11 | 15 | 15 |
| 172 | CURR EYE RES | 3 | 28 | 32 | 24 | 24 | 30 | 20 |
| 173 | CURR OPIN OPHTHALMOL | 3 | 14 | 14 | 19 | 16 | 9 | 4 |
| 174 | CUTAN OCUL TOXICOL | 3 | 46 | 49 | 52 | 49 | 41 | 43 |
| 175 | DOC OPHTHALMOL | 3 | 26 | 44 | 35 | 49 | 32 | 34 |
| 176 | EUR J OPHTHALMOL | 3 | 45 | 46 | 27 | 31 | 42 | 47 |
| 177 | EXP EYE RES | 3 | 8 | 9 | 9 | 11 | 6 | 3 |
| 178 | EYE | 3 | 23 | 26 | 14 | 14 | 27 | 30 |
| 179 | EYE CONTACT LENS | 3 | 24 | 30 | 27 | 31 | 16 | 28 |
| 180 | GRAEF ARCH CLIN EXP | 3 | 20 | 20 | 16 | 15 | 22 | 20 |
| 181 | INDIAN J OPHTHALMOL | 3 | 47 | 48 | 4 | 5 | 18 | 17 |
| 182 | INT J OPHTHALMOL-CHI | 3 | 54 | 54 | 56 | 52 | 55 | 53 |
| 183 | INVEST OPHTH VIS SCI | 3 | 6 | 6 | 2 | 2 | 3 | 2 |
| 184 | JAMA OPHTHALMOL | 3 | | | 49 | 36 | 12 | 26 |
| 185 | JPN J OPHTHALMOL | 3 | 31 | 29 | 27 | 26 | 37 | 39 |
| 186 | J FR OPHTALMOL | 3 | 53 | 56 | 54 | 55 | 55 | 57 |
| 187 | J AAPOS | 3 | 42 | 43 | 25 | 36 | 45 | 43 |
| 188 | J CATARACT REFR SURG | 3 | 12 | 15 | 6 | 5 | 14 | 13 |
| 189 | J EYE MOVEMENT RES | 3 | 37 | 47 | 57 | 57 | 55 | 51 |
| 190 | J GLAUCOMA | 3 | 19 | 17 | 22 | 22 | 25 | 25 |
| 191 | J NEURO-OPHTHALMOL | 3 | 34 | 28 | 35 | 24 | 21 | 42 |
| 192 | J OCUL PHARMACOL TH | 3 | 35 | 34 | 27 | 36 | 29 | 29 |
| 193 | J OPHTHALMOL | 3 | 25 | 25 | 49 | 40 | 50 | 36 |
| 194 | J PEDIAT OPHTH STRAB | 3 | 49 | 51 | 54 | 55 | 47 | 52 |
| 195 | J REFRACT SURG | 3 | 13 | 11 | 13 | 11 | 7 | 9 |
| 196 | J VISION | 3 | 9 | 12 | 12 | 18 | 37 | 14 |
| 197 | KLIN MONATSBL AUGENH | 3 | 51 | 53 | 49 | 45 | 47 | 54 |
| 198 | MOL VIS | 3 | 17 | 21 | 17 | 16 | 22 | 6 |
| 199 | OCUL IMMUNOL INFLAMM | 3 | 39 | 33 | 27 | 31 | 35 | 37 |
| 200 | OCUL SURF | 3 | 5 | 4 | 27 | 26 | 8 | 17 |
| 201 | OPHTHAL PHYSL OPT | 3 | 29 | 13 | 25 | 22 | 12 | 16 |
| 202 | OPHTHAL EPIDEMIOL | 3 | 27 | 39 | 35 | 40 | 36 | 23 |
| 203 | OPHTHALMIC GENET | 3 | 43 | 41 | 46 | 49 | 55 | 33 |
| 204 | OPHTHAL PLAST RECONS | 3 | 48 | 50 | 43 | 36 | 46 | 48 |
| 205 | OPHTHALMIC RES | 3 | 40 | 35 | 35 | 31 | 34 | 31 |
| 206 | OPHTHAL SURG LAS IM | 3 | 44 | 37 | 43 | 45 | 52 | 45 |
| 207 | OPHTHALMOLOGE | 3 | 50 | 52 | 43 | 45 | 54 | 50 |
| 208 | OPHTHALMOLOGICA | 3 | 36 | 27 | 27 | 26 | 31 | 32 |
| 209 | OPHTHALMOLOGY | 3 | 2 | 2 | 1 | 1 | 2 | 5 |
| 210 | OPTOMETRY VISION SCI | 3 | 18 | 22 | 19 | 21 | 27 | 22 |
| 211 | OPTOMETRY | 3 | 38 | 36 | 46 | 52 | 20 | 55 |
| 212 | PROG RETIN EYE RES | 3 | 1 | 1 | 6 | 9 | 1 | 1 |
| 213 | RETINA-J RET VIT DIS | 3 | 10 | 8 | 10 | 3 | 10 | 11 |
| 214 | REV BRAS OFTALMOL | 3 | 55 | 57 | 57 | 57 | 50 | 58 |
| 215 | SEMIN OPHTHALMOL | 3 | | 42 | 35 | 40 | 44 | 34 |
| 216 | SURV OPHTHALMOL | 3 | 7 | 7 | 17 | 18 | 4 | 12 |
| 217 | VISION RES | 3 | 15 | 18 | 10 | 10 | 24 | 8 |
| 218 | VISUAL NEUROSCI | 3 | 30 | 31 | 41 | 31 | 17 | 24 |
| 219 | ADV ENERGY MATER | 4 | 4 | 3 | 7 | 6 | 1 | 1 |
| 220 | ADV FUNCT MATER | 4 | 5 | 5 | 5 | 5 | 5 | 4 |
| 221 | ADV MATER | 4 | 2 | 2 | 1 | 1 | 2 | 3 |
| 222 | ADV COND MATTER PHYS | 4 | 41 | 44 | 50 | 53 | 45 | 44 |
| 223 | APPL SURF SCI | 4 | 11 | 13 | 9 | 8 | 12 | 10 |
| 224 | CHEM VAPOR DEPOS | 4 | 23 | 37 | 42 | 45 | 30 | 33 |
| 225 | CONDENS MATTER PHYS | 4 | 47 | 48 | 51 | 55 | 51 | 52 |
| 226 | EUR PHYS J B | 4 | 31 | 35 | 21 | 29 | 43 | 36 |
| 227 | FERROELECTRICS | 4 | 53 | 55 | 46 | 50 | 51 | 55 |
| 228 | FERROELECTRICS LETT | 4 | 51 | 53 | 58 | 56 | 51 | 51 |
| 229 | IEEE T SEMICONDUCT M | 4 | 42 | 45 | 48 | 45 | 46 | 46 |
| 230 | INTEGR FERROELECTR | 4 | 56 | 56 | 51 | 53 | 51 | 57 |
| 231 | INT J MOD PHYS B | 4 | 57 | 54 | 42 | 40 | 38 | 54 |
| 232 | IONICS | 4 | 26 | 26 | 35 | 36 | 31 | 27 |
| 233 | J COMPUT THEOR NANOS | 4 | 45 | 43 | 35 | 34 | 32 | 43 |
| 234 | J LOW TEMP PHYS | 4 | 44 | 42 | 38 | 37 | 50 | 41 |
| 235 | J MAGN | 4 | | 58 | 56 | 56 | 51 | 58 |
| 236 | J MAGN MAGN MATER | 4 | 19 | 20 | 11 | 13 | 24 | 21 |
| 237 | J MATER SCI-MATER EL | 4 | 29 | 22 | 33 | 29 | 24 | 26 |
| 238 | J NANOSCI NANOTECHNO | 4 | 38 | 38 | 17 | 12 | 32 | 39 |
| 239 | J PHYS CHEM SOLIDS | 4 | 28 | 32 | 28 | 29 | 28 | 22 |
| 240 | J PHYS-CONDENS MAT | 4 | 14 | 15 | 8 | 9 | 16 | 18 |
| 241 | J SUPERCOND NOV MAGN | 4 | 46 | 46 | 38 | 34 | 38 | 45 |
| 242 | LASER PHOTONICS REV | 4 | 6 | 6 | 11 | 10 | 7 | 7 |
| 243 | MATER SCI ENG B-ADV | 4 | 22 | 18 | 25 | 18 | 26 | 19 |
| 244 | MAT SCI SEMICON PROC | 4 | 35 | 27 | 46 | 37 | 22 | 16 |
| 245 | METALLOFIZ NOV TEKH+ | 4 | 59 | 60 | 60 | 59 | 51 | 59 |
| 246 | MOD PHYS LETT B | 4 | 52 | 50 | 41 | 45 | 47 | 50 |
| 247 | NANO LETT | 4 | 3 | 4 | 2 | 2 | 3 | 2 |
| 248 | NAT MATER | 4 | 1 | 1 | 3 | 3 | 4 | 9 |
| 249 | PHASE TRANSIT | 4 | 43 | 41 | 44 | 40 | 35 | 42 |
| 250 | PHILOS MAG | 4 | 30 | 36 | 32 | 33 | 42 | 34 |
| 251 | PHIL MAG LETT | 4 | 39 | 40 | 38 | 44 | 35 | 37 |
| 252 | PHYSICA B | 4 | 40 | 39 | 24 | 23 | 38 | 40 |
| 253 | PHYSICA E | 4 | 34 | 25 | 14 | 16 | 21 | 30 |
| 254 | PHYS STATUS SOLIDI A | 4 | 33 | 33 | 21 | 24 | 41 | 35 |
| 255 | PHYS STATUS SOLIDI B | 4 | 36 | 31 | 20 | 16 | 37 | 38 |
| 256 | PHYS STATUS SOLIDI-R | 4 | 17 | 14 | 28 | 24 | 11 | 14 |
| 257 | PHYS REV B | 4 | 9 | 9 | 4 | 4 | 9 | 8 |
| 258 | PHYS CHEM LIQ | 4 | 50 | 52 | 51 | 49 | 44 | 48 |
| 259 | PHYS SOLID STATE+ | 4 | 48 | 47 | 44 | 40 | 48 | 47 |
| 260 | PLASMA PROCESS POLYM | 4 | 10 | 10 | 25 | 27 | 15 | 15 |
| 261 | PROG SURF SCI | 4 | 8 | 8 | 34 | 40 | 8 | 6 |
| 262 | RADIAT EFF DEFECT S | 4 | 54 | 51 | 51 | 50 | 51 | 53 |
| 263 | SEMICOND SCI TECH | 4 | 24 | 17 | 21 | 18 | 19 | 31 |
| 264 | SEMICONDUCTORS+ | 4 | 49 | 49 | 48 | 45 | 48 | 49 |
| 265 | SEMICONDUCT SEMIMET | 4 | 37 | 28 | 56 | 50 | 51 | 31 |
| 266 | SMALL | 4 | 7 | 7 | 6 | 7 | 6 | 5 |
| 267 | SOLID STATE COMMUN | 4 | 25 | 29 | 13 | 14 | 16 | 28 |
| 268 | SOLID STATE ELECTRON | 4 | 32 | 34 | 30 | 29 | 26 | 29 |
| 269 | SOLID STATE IONICS | 4 | 13 | 19 | 14 | 14 | 18 | 11 |
| 270 | SOLID STATE NUCL MAG | 4 | 15 | 11 | 35 | 37 | 10 | 17 |
| 271 | SOLID STATE SCI | 4 | 20 | 30 | 25 | 27 | 22 | 19 |
| 272 | SOLID STATE TECHNOL | 4 | 58 | 59 | 58 | 60 | 51 | 60 |
| 273 | SUPERCOND SCI TECH | 4 | 12 | 12 | 14 | 18 | 14 | 13 |
| 274 | SUPERLATTICE MICROST | 4 | 27 | 21 | 30 | 24 | 13 | 25 |
| 275 | SURF REV LETT | 4 | 55 | 57 | 51 | 56 | 51 | 56 |
| 276 | SURF SCI | 4 | 21 | 23 | 18 | 18 | 20 | 23 |
| 277 | SYNTHETIC MET | 4 | 16 | 16 | 18 | 18 | 29 | 12 |
| 278 | THIN SOLID FILMS | 4 | 18 | 24 | 9 | 11 | 34 | 24 |

Source: Thomson Reuters Web of Science. WOS Categories: 1 = INFORMATION SCIENCE & LIBRARY SCIENCE; 2 = OPERATIONS RESEARCH & MANAGEMENT SCIENCE; 3 = OPHTHALMOLOGY; 4 = PHYSICS, CONDENSED MATTER.